\documentclass[sigconf]{acmart}
\renewcommand\footnotetextcopyrightpermission[1]{} % removes footnote with conference information in first column

\usepackage{multirow}
\usepackage{booktabs}
\usepackage{graphicx}
\usepackage{caption}
\usepackage{subcaption}
\usepackage{enumerate}

\usepackage[ruled, linesnumbered]{algorithm2e}
\renewcommand{\KwOut}{\textbf{Repeat:}}
\renewcommand{\KwResult}{\textbf{Until:}}

\newlength\savewidth

\AtBeginDocument{%
  \providecommand\BibTeX{{%
    \normalfont B\kern-0.5em{\scshape i\kern-0.25em b}\kern-0.8em\TeX}}}

\begin{document}
\fancyhead{}
% \title{PCRec: Enhancing recommendation systems by multiple models self-cooperation}
\title{Enhancing Top-N Item Recommendations by Peer Collaboration}


\author{Yang Sun}
% \authornote{Joint first author.}
\affiliation{%
  \institution{University of Science and Technology of China}
 }
\email{yang.sun@siat.ac.cn}

\author{Fajie Yuan}
\authornote{Joint first author.}
\affiliation{%
  \institution{Westlake University}}
\email{yuanfajie@westlake.edu.cn}

\author{Min Yang}
\authornote{Min Yang is corresponding author. This work was conducted when Yang Sun was interning at SIAT, Chinese Academy of Sciences.}
\affiliation{%
  \institution{SIAT, Chinese Academy of Sciences}}
\email{min.yang@siat.ac.cn}

% \cortext[cor1]{Corresponding author}

\author{Alexandros Karatzoglou}
\affiliation{%
  \institution{Google}}
\email{alexandros.karatzoglou@gmail.com}

\author{Li Shen}
\affiliation{%
  \institution{JD Explore Academy}}
\email{mathshenli@gmail.com}

\author{Xiaoyan Zhao}
\affiliation{%
  \institution{SIAT, Chinese Academy of Sciences}}
\email{xiaoyan.zhao@siat.ac.cn}

% \author{Duo Liu}
% \affiliation{%
%   \institution{University of Science and Technology of China}}
% \email{duo.liu@siat.ac.cn}


\begin{abstract}
Deep neural networks (DNN) have achieved great success in the recommender systems (RS) domain. However, to achieve remarkable performance, DNN-based recommender models often require numerous parameters, which inevitably bring redundant neurons and weights, a phenomenon referred to as over-parameterization.
In this paper, we plan to exploit such redundancy phenomena to improve the performance of RS. Specifically, we propose PCRec, a top-N item \underline{rec}ommendation framework that leverages collaborative training of two DNN-based recommender models with the same network structure, termed \underline{p}eer \underline{c}ollaboration.
PCRec can reactivate and strengthen the unimportant (redundant) weights during training, which achieves higher prediction accuracy but maintains its original inference efficiency. To realize this, we first introduce two  criterions to identify the importance of weights of a given recommender model. Then, we rejuvenate the unimportant weights by transplanting outside information (i.e., weights) from its peer network. After such an operation and retraining, the original recommender model is endowed with more representation capacity by possessing more functional model parameters. To show its generality, we instantiate PCRec by using three well-known recommender models. We conduct extensive experiments on three real-world datasets, and show that PCRec yields significantly better recommendations than its counterpart with the same model (parameter) size. 
% Our code is available at \url{https://github.com/anonymouskdd2021/PeerCollaboration}.
% , with the same network structure, that is optimized in a standard way. 

% Deep neural networks achieve the state-of-the-art result in recommendation system. However, existing a large number of unimportant weights in such models. A series of methods can effectively improve the efficiency of the model by removing redundant weights such as pruning, but meanwhile, drop the potential of the model. To more effectively utilize these unimportant weights, we do the opposite and propose a multiple models \underline{se}lf-\underline{co}operation framework, terms as PCRec, where external information of peers of each model is complemented to the unimportant weights of the model to make them play a more important role in developing the potential of the model, and significantly improve the model performance. We instantiate PCRec using three powerful deep neural networks. which is a deep convolutional neural network with dilated kernels which is state-of-the-art result sequential recommendation system, neural factorization machine, and YouTubeDNN of non-sequential recommendation system, given consideration to both recommendation accuracy and efficiency, and further verify the adaptability of PCRec. By the extensive ablation studies, we demonstrate that the proposed PCRec can achieve compelling results in real-world RS datasets. Meanwhile, PCRec only needs one member during inference, which differs from other collaborative learning of multiple models such as ensemble learning, and PCRec always outperforms these methods.
\end{abstract}

\maketitle

\section{INTRODUCTION}
\label{intro}
Recommender Systems (RS) have become an essential tool for large social media and e-commerce platforms. A large number of user-item interaction behaviors (i.e., feedback) are produced explicitly or implicitly every day on such systems \cite{yuan2020future}. In particular, implicit feedback, such as clicks, purchases, watched videos and played songs, are easy to be collected and often at a very large scale.
% much easier to be collected and at a larger scale than explicit ratings.
For example, users on Tiktok may easily watch thousands of short videos per day, given that the playing time of each video takes usually less than 20 seconds. As such, recent studies on top-N item recommendations mainly pay attention to the implicit feedback problem~\cite{bayer2017generic}.
The essence of item recommendation from implicit feedback is to predict a list of top-N items that a user would like to interact with by learning from his/her previous feedback. 

\begin{figure}[t]
	\centering
	\begin{subfigure}[t]{0.25\textwidth}
		\centering
		\includegraphics[width=1.7in]{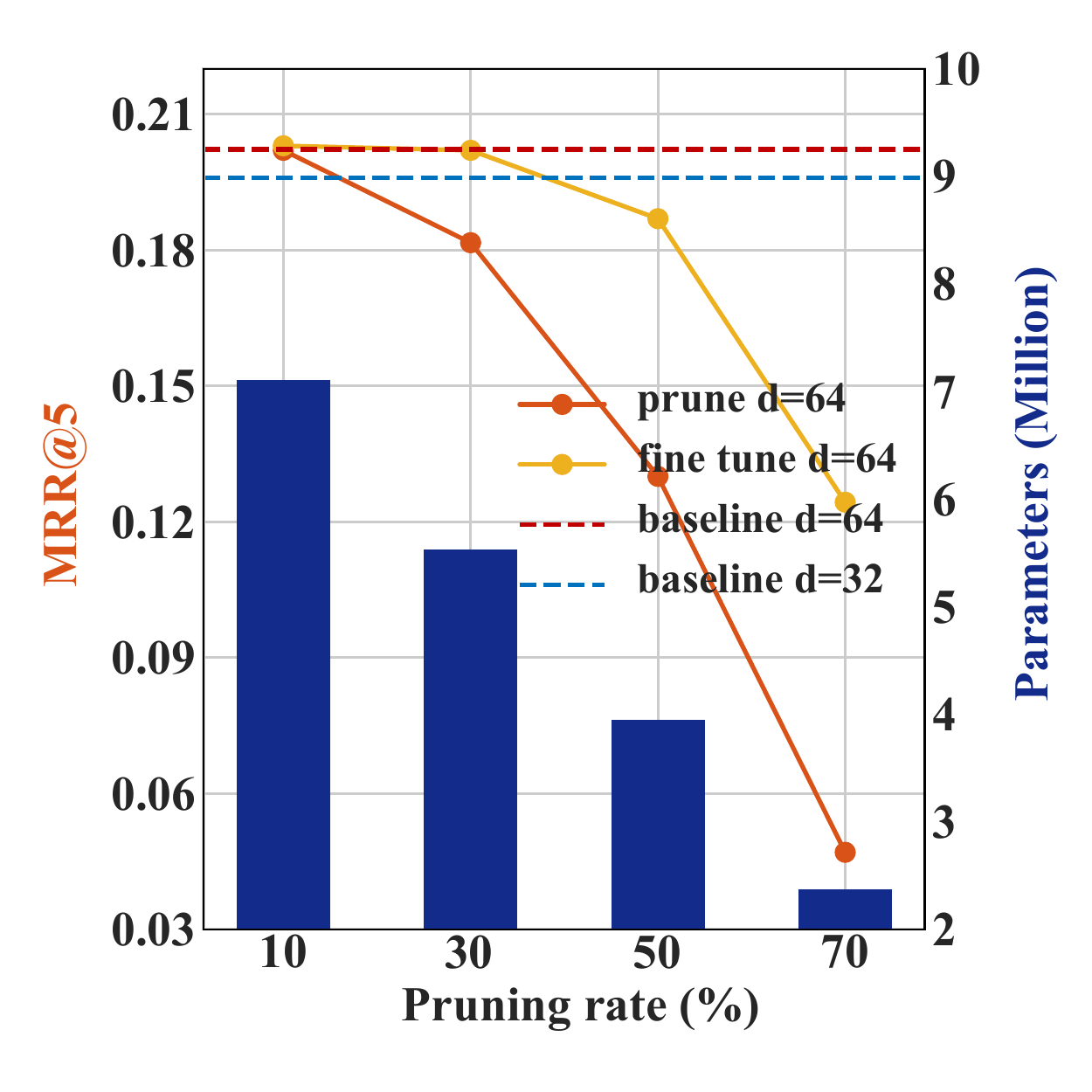}
		\subcaption{Retailrocket}
		\label{fig:deep}
	\end{subfigure}%
	\begin{subfigure}[t]{0.25\textwidth}
		\centering
		\includegraphics[width=1.7in]{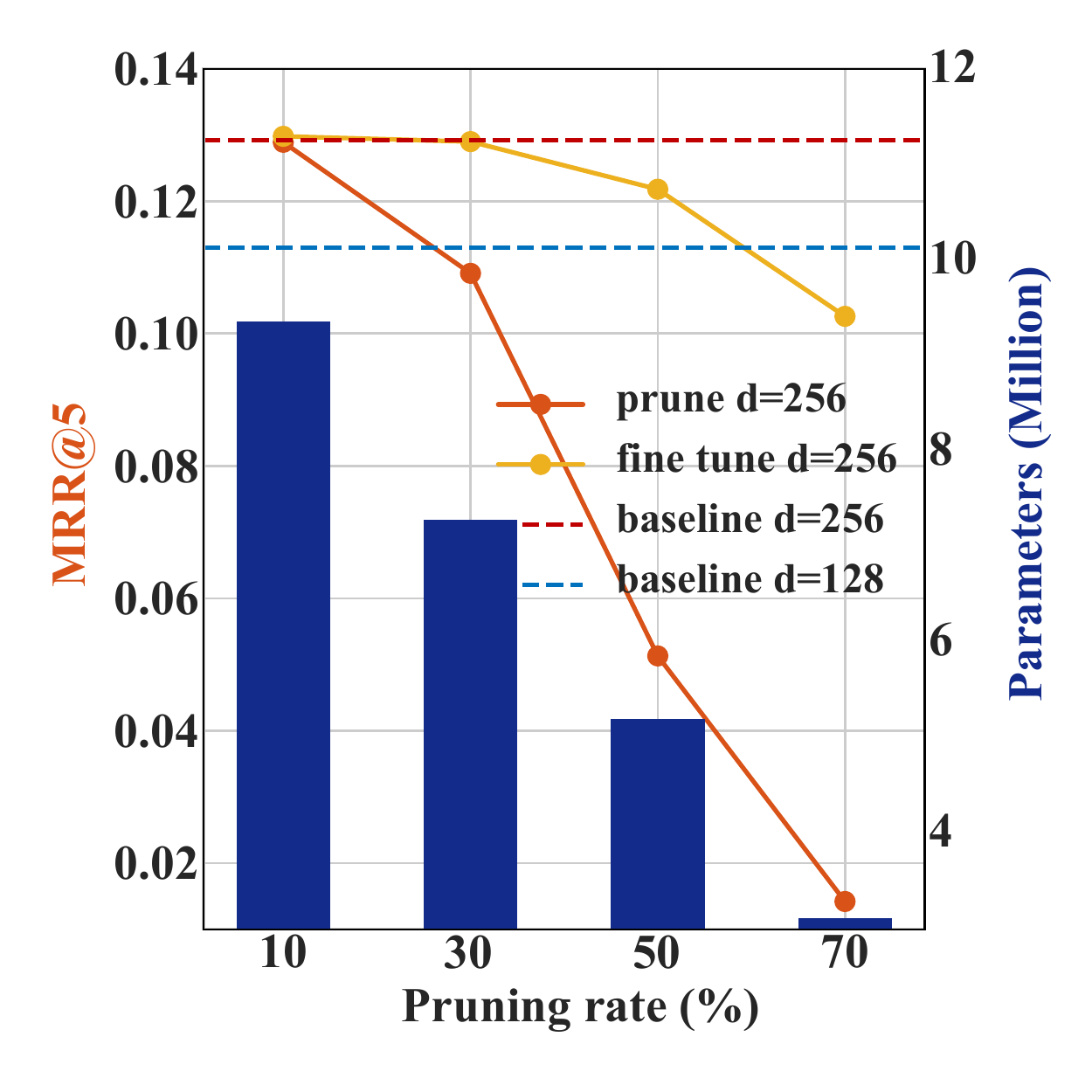}
		\subcaption{ML-20M}
		\label{fig:wide}
	\end{subfigure}
	\caption{\small Performance change by pruning on SASRec. We perform the standard pruning based on the weight magnitude following~\cite{see2016compression}. The experimental settings are given in Section~\ref{expsetup}.}
	\label{prunefigure}
	\vspace{-0.1in}
\end{figure}
% \begin{figure}[t]
%   \centering
%   \includegraphics[width=0.45\textwidth]{prune_neural2.pdf}
%   \caption{Multi-model cooperation framework, where we take four selfsame networks as an example.}
%   \label{multiplemodels}
% \end{figure}
Embedding and deep neural networks (DNN) based recommender models have achieved superior performance and practically dominated the RS domain. Among these models, BPR \cite{rendle2012bpr}, DSSM~\cite{huang2013learning} and  YouTube DNN~\cite{covington2016deep} have become some of the most representative work for the general item recommendation task, while
GRU4Rec~\cite{hidasi2015session}, NextItNet~\cite{yuan2019simple} and SASRec~\cite{kang2018self} are more representative for the sequential recommendation settings. The success of these models often comes with a large  embedding size  or deep network structure~\cite{sun2020generic,wang2020stackrec}. 
% Taking NextItNet as an example, its highest recommendation accuracy is obtained with up to 30 convolutional layers on the benchmark movielens dataset~\cite{sun2020generic}.
However, large and deep models are very prone to be over-parameterized, resulting in redundant neurons and weights\footnote{We use weights and parameters interchangeably in this paper. Unimportant or redundant weights are the weights that have no effect on model expressivity}. As illustrated in Figure~\ref{prunefigure}, simply pruning 10\% parameters in the SASRec model yields very minor performance degradation. What's more, pruning 30\% of unimportant parameters with a fine-tuning strategy performs even a bit better than the original SASRec. On the other hand, SASRec with a smaller embedding dimension (i.e., $d=128$), around 50\% parameters of itself with $d=256$, performs noticeably worse on ML-20M. These observations evidence that (1) the over-parameterization phenomenon widely exists in large recommender models; (2) training a smaller-size recommender model from scratch yields considerably worse performance.
Pruning redundant parameters from a large neural network model could bring higher parameter efficiency.\footnote{Note that fewer parameters does not necessarily lead to efficient training or inference.} These experiments have been  extensively performed in the computer vision (CV)~\cite{han2015learning,han2015deep,frankle2018lottery} and natural language processing (NLP)~\cite{lan2019albert,gordon2020compressing} fields. However, in recommender systems, simply reducing a portion of parameters (e.g., 30\% in Figure~\ref{prunefigure}) for large recommender models may not benefit as much as in CV and NLP since large-scale RS models are often deployed in a cloud platform rather than an edge/mobile device (like many CV and NLP models) with very limited hardware resources. Thereby, inspired by these work, but different from them, we hope to explore whether such redundant parameters can be used more effectively instead of abandoning them  so as to increase the model expressivity 
and alleviate the data sparsity issue in the recommender system domain.

% we yet argue that a large network with more valid parameters could be more expressive and performs better. In view of this, in this paper, we explore how to exploit such redundant parameters and reactivate them so as to further enhance the expressivity of deep recommender model.
% and capacity 
% a larger capacity and potential generally as the presentation in \cite{sun2020generic}, 
%  Intuitively,
%  we can strengthen these invalid weights
%by transplanting some important weights from one or more identical model(s).
% further develop the potential of recommender models
% simply removing parameters may reduce the potential of the model. Motivated by these empirical findings, whether we can utilize these unimportant weights and revise them to further develop the potential of recommender models, which enable better model generalization ability.
% , which means we need to consider these important weights that have the higher status than these unimportant weights. 
To approach the above problem, we present a \underline{p}eer \underline{c}ollaboration framework for top-N item \underline{rec}ommendation tasks, called PCRec. Specifically, we propose rejuvenating invalid (i.e., unimportant) weights of a recommender model by transplanting important weights from a peer model with an identical\footnote{`Identical' here only means the same network architecture, rather than their parameters and hyper-parameters throughout this paper.} network.
% To do so,  PCRec solves this issue from  the layer-wise perspective rather than weight-wise
% focus on the informative importance of layer rather than directly operating on model weights.
%  we complement invalid weights of each model by external information stem from other models with diversity parameters and the same architecture. 
To do so, we first propose two criteria, including L1-norm based and entropy based, to identify which weights are important and which are redundant. 
% Note that \replaced{L1-norm-based criterion}{an alternative is to measure the importance of weights by L1-norm or L2-norm,  which} are widely used for network pruning~\cite{see2016compression}. However, compared with the entropy-based criterion, L1-norms care only about the absolute values of weights, and fail to measure the variation of weights \cite{You2020GraphSO, blundell2015weight, paninski2003estimation}.
%  First, we focus on the layer as measure unit between multiple models, rather than just the unimportant weights that are discretely distributed across each dimension in each layer of the model. Because each layer of model integrally has different effects for a specific task and the parameters in a layer interact with each other \cite{zeiler2014visualizing, jawahar2019does, yuan2019simple}. Second, we determine whether the weights (layers) are important or unimportant by the criterion, entropy which can precisely indicate the information capacity of the measure unit, rather than L1-norm or L2-norm which widely were used for network pruning due to they only consider the value of weights and omit the relationship between weights.  level of "information"
To effectively strengthen invalid weights, we create two rules regarding how to complement information between two identical networks and how much information each one needs to be complemented from its peer. To validate the efficacy of PCRec, we instantiate it using three popular models, including  both general item recommender models and sequential recommender models. 

% For the general item recommendations, we integrate PCRec with BPR and YouTubeDNN, which are well-known in both industrial and academical communities~\cite{rendle2012bpr,covington2016deep}. As for sequential recommendations, we instantiate PCRec using SASRec, which is the state-of-the-art baseline for modeling user sequence data~\cite{kang2018self}.

% The first is NextItNet-style sequential recommendation system as the baseline considering that it achieves state-of-the-art performance in modeling sequential dependency of user-item interactions. Besides, we also use NeuralFM model based on factorization method as another baseline because it produces start-of-the-art results by modeling higher-order and non-linear feature interactions for sparse data prediction. The last one is YouTubeDNN that is simple yet effective model to capture strong recommendation accuracy by modeling non-linear interactions between item latent features. We name the proposed framework PCRec, where ‘SeCo’ stands for self-cooperation of multiple models with the same architecture.

We summarize our main contributions as four-fold:
\begin{itemize}
\item We propose PCRec to promote collaboration of two recommender models with a selfsame network architecture. PCRec is a novel learning paradigm for recommender models, which can
reactivate invalid weights by \textit{explicitly} transplanting effective weights from its outside peer network.
% To the best of our knowledge, PCRec is the first recommender framework that 
% to complement information for each layer of every model that have unimportant weights circularly and improve the generalization of these models, and the performance is better as the number of models increases.
\item We introduce two criteria to measure the importance of weights in a recommender model. Besides, we propose an adaptive coefficient to determine how much the external information is required from its peer.
% , and put forward a linear combination formulation to form their cooperation.
\item We instantiate PCRec using three well-known recommender models, namely, BPR, YouTube DNN, and SASRec. PCRec is conceptually simple, easy to implement, and applicable to a broad class of recommender models. 
\item Through thorough experiments and ablation studies, we show that PCRec obtains noticeably improved performance on three real-world RS datasets. 
% For reproducibility, we submit the code and data anonymously: \textcolor{blue}{https://github.com/anonymouskdd2021/PeerCollaboration}.%Our code is available at \textcolor{blue}{https://github.com/anonymouskdd2021/PeerCollaboration}.
\end{itemize}

\section{RELATED WORK}
We briefly review related work regarding the DNN-based RS and multiple model ensemble learning.

\subsection{Item Recommendation with Deep Learning}
Deep neural networks (DNNs) have made great progress for item recommendations thanks to their high model capacity and expressivity. In general, deep RS can be broadly classified into general (i.e., non-sequential) item recommendations and sequential item recommendations according to whether sequential patterns are modeled.  In terms of general item recommendations, neural network models such as  Deep Crossing~\cite{shan2016deep}, DeepFM~\cite{guo2017deepfm}, NeuralFM~\cite{he2017neural}, Wide \& Deep~\cite{cheng2016wide}, and YouTube DNN~\cite{covington2016deep} have become the most representative works. 
Compared to the shallow embedding models, the main advantages of these models highly depend on their neural network structures and non-linearities,  who are believed to be able to approximate any continuous function~\cite{hornik1991approximation, hornik1989multilayer}. 

On the other hand, sequential recommender systems (SRS) have also attracted much attention recently. By capturing user's dynamic interests, SRS, in general, is more powerful in generating the next recommendation. Moreover, SRS can be trained in a self-supervised manner~\cite{yuan2020parameter,zhou2020s3}, and thus do not need handcrafted labels and features. According to existing literature, GRU4Rec~\cite{hidasi2015session}, Caser~\cite{tang2018personalized}, NextItNet~\cite{yuan2019simple}, SASRec~\cite{kang2018self} and BERT4Rec~\cite{sun2019bert4rec} are especially popular. Among them, GRU4Rec and Caser based on shallow network structure fail to model very long-term sequential patterns and usually offer sub-optimal performance. By contrast, NextItNet, SASRec and BERT4Rec are able to obtain state-of-the-art performance by effectively capturing long-term and complex sequential dependencies.
% due to the effective design of the residual units.
% {GRU4Rec and Caser are relatively shallower models since stacking more hidden layers does not lead to improved results. By contrast, recent work in~\cite{sun2020generic} demonstrated that NextItNet could be stacked with up to 30 convolutional layers due to its effective design of  the residual units.}

In this paper, we design PCRec by instantiating it with three popular recommender models including BPR with shallow embeddings, YouTube DNN and SASRec with deep neural network. It is worth noting that the framework of PCRec is model-agnostic and potentially applicable to various embedding and deep models.

\subsection{Multiple Model Learning}
PCRec relates to the ensemble learning (EL) \cite{hansen1990neural, krogh1995neural} and knowledge distillation (KD) \cite{hinton2015distilling} in a similar spirit that more than one model is used during training. Here, we briefly review related works and clarify their key differences against PCRec.
% We concisely summarize ensemble learning and knowledge distillation into multi-model learning.
% \subsubsection{Ensemble learning}

Ensemble Learning (EL) refers to the process that multiple learning models are strategically combined to achieve better predictive performance than any of its individual model trained alone~\cite{opitz1999popular}. Bagging~\cite{breiman1996bagging}, boosting~\cite{friedman2002stochastic} and stacking~\cite{sigletos2005combining} are thought of as three representative EL algorithms. The main principle behind them is that a set of weak learners are combined together to form a strong learner. While EL is generic for different types of models, we notice that there are relatively few works that explore deep learning (DL) based ensemble methods. We suspect that DL-based methods are not conceptually weak learners and combining a large number of DL models could be computationally and memory expensive during the model prediction phase, and thus could be in-efficient and in-practical. By contrast, PCRec merely needs one well-trained single model at the inference stage.  
% Meanwhile, unlike the standard ensemble methods that are designed to reduce bias or

Apart from that, PCRec is also relevant to KD-based methods~\cite{hinton2015distilling,tang2018ranking} which are designed to enhance a small-capacity model by one (or multiple) large teacher model(s). 
However, unlike KD-based methods, PCRec does not include the mutual learning~\cite{zhang2018deep,liu2019daml} process which optimizes multiple losses together. Moreover, PCRec explicitly combines the advantage of two identical models by enhancing the invalid weights, which is very explainable. By contrast, the knowledge transferred by KD-based methods is usually called dark knowledge~\cite{hinton2015distilling}, and the working mechanism of it is not as explainable as PCRec. In addition,  PCRec focuses on performance improvement which is different 
from  the motivation of the KD-based methods --- injecting knowledge from a large teacher model into a smaller student one to obtain the effect of model compression.

\section{METHODS}
As mentioned in the introduction part, over-parameterization or redundancy commonly exist in large and deep recommender models. Inspired by this, in this paper we set our goal to reactivate these redundant weights (rather than abandoning them) so as to enhance the model capacity and expressivity.  

To be specific, we present the PCRec learning framework, which enhances an individual recommender model by transplanting important information from a selfsame network of this recommender, referred to as a peer. In the following, we first introduce criteria to measure the importance of weights in a recommender model. 
% we determine the layer of model as  basic unit to measure the importance of weights (i.e. information capacity). 
% Second, we adapt a criterion for calculating the information capacity of the layer. 
Then, we propose a parameter-wise approach to reactivate the redundant weights of the two peer models. At last, we  develop the final version based on the layer-wise cooperation, which 
 addresses the limitations of the parameter-wise approach. 
\begin{figure}[t]
  \centering
  \includegraphics[width=0.45\textwidth]{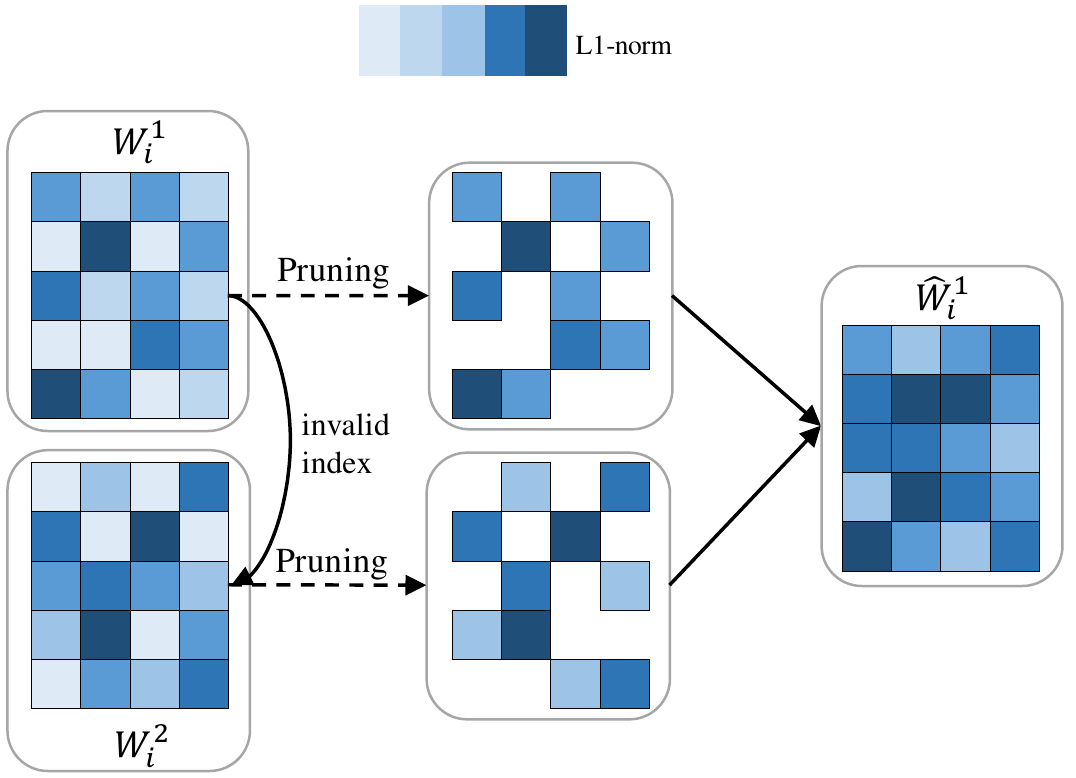}
  \caption{PCRec with PW cooperation, where dark colors represent important weight.}
  \label{individualweight}
\end{figure}

\subsection{{Criterion}}
\label{citeria}

\subsubsection{L1-norm}
The idea of the L1-norm criterion is borrowed from  the pruning~\cite{han2015deep, see2016compression, li2016pruning} literature.
% , where important weights means
% the one that has a higher absolute value, otherwise is an invalid weight. 
% \added{Inspired by previous works~\cite{han2015deep, see2016compression, li2016pruning} that use  L1-norm based criterion (i.e., magnitude of weights) for network pruning, we
% want to use the same criterion for information transplanting.
% we adopt L1-norm based criterion like weight pruning, which consists of pruning weights with smallest absolute value, to measure the magnitude of weights.
Denote $W_{i} \in  \mathbb{R}^{d_{i-1} \times d_{i}}$ as the weight matrices of the $i$-th layer in a model. We can identify the importance of weights from two perspectives: single weight perspective and entire layer perspective.
% \added{correspondingly, we can operate information transplanting (mentioned later) from the two perspectives.} 
The importance of a single weight is directly determined by its absolute value (L1-norm) --- the higher absolute value it has, the more important it is. We can use a threshold to distinguish the important and unimportant weights.
% \added{which we only activate these unimportant weights}.
From the entire layer perspective, we could identify the importance of all weights by using a neural network layer (including the embedding, middle layers, and final prediction layers) as the measure unit. Intuitively, measuring the importance of a layer could maintain the layer consistency as much as possible, which will benefit the information transplanting process as mentioned later. Formally, its L1-norm can be given below by using the entire layer as the measure unit:
\begin{equation}
\| W_{i}||=\sum_{m=1}^{d_{i-1}} \sum_{n=1}^{d_{i}}|W_{i, m, n}|
\end{equation}
% \deleted{where  $W_{i}$  is the $i$-th layer with shape}\footnote{For simplicity, we illustrate PCRec by using the 2D weight matrix as an example, although weights in this paper can be any dimensions.} $m\times n$. 
Denote $W_{i}^1$ and $W_{i}^2$  as L1-norm of $i$-th layer of two collaborated recommender models.
We define the $H(W_{i}^k)$ as the relative information of the layer:
\begin{equation}
\label{indiweight}
\mathrm{H}(W_{i}^{k})=\frac{\|W_{i}^{k}\|}{\|W_{i}^{k}\|+\|W_{i}^{k\%2+1}\|}, \quad k=1,2.
\end{equation}
where $\%$ is the modulo operation, $k$ is the model ID. 
While L1-norms has been widely applied in pruning, it only cares about the magnitude of the weights, and ignores the variation of the weights. 
% i.e., the quantity of information.
For example, given a weight matrix  $W_{i} \in R^{d_{i-1} \times d_{i}}$ of $i$-th layer, where each element in $W_{i}$ is assigned to the same value $z$, whose absolute value is big. If we use such a weight matrix to transform the $(i-1)$-th layer, then each part of it contributes equally to the $i$-th layer even $z$ is very big. This suggests that L1-norm might not be the best criterion to discriminate the importance of layer (all weights) information.
\begin{algorithm}[t]

    \caption{Parameter-wise Cooperation}
    \label{pw_alg}
    % \Indm\Indmm
    \KwIn{
    two models ${M}_1, {M}_2$, where each one has $N$ layers with weights ${W}_1^k, {W}_2^k, \dots, {W}_N^k$, $k = 1, 2$; two different shuffles for train data: $\mathcal{D}_1, \mathcal{D}_2$;
   two different learning rates: $\eta_1, \eta_2$; the threshold $\gamma$;}
    $t = 0.$ \\
    \SetAlgoNoLine
    \KwOut{ \\
        \Indp\Indpp
        $t = t + 1$ \\
        \For{k = 1, 2}{
            Compute the gradient and Update $M_k$ with $\eta_k$ and $\mathcal{D}_k$; \\
            \tcc{Update parameters at each batch for all models parallelly or serially.}
        }
    
        \For{k = 1, 2}{
            \For{i = 1, 2, \dots, N}{
                Calculate index $I_i^k$  with $\gamma$ by Eq.~(\ref{index}) \;
                Get $\widehat{W}_i^k$
                % $\widehat{W}_i^k = W_{i}^k * I_i^k + W_{i}^{k+1} * (1-I_i^k)$ 
                by Eq.~(\ref{pw-newweight}) ; 
                \tcc{Replace $ W_i^k$ with $\widehat{W}_i^k$ in $M_k$.}
            }
        }
    % \Repeat{}{}
    }
    \KwResult{convergence}  
\end{algorithm}

\subsubsection{Entropy}
% \deleted{Instead of measuring the information of a single weight, we choose all weights in a neural network layer (including the embedding and softmax layers) as the measure unit. The main purpose here is to maintain the network connectivity as much as possible when rejuvenating the invalid weights.}
% When we determine the layer as a measure unit, L1/L2-norms only calculate the magnitude of weights in layer, which is not fully represent the information and variant of weights in the layer. Given weight matrix $W_{i} \in R^{d_{i-1} \times d_{i}}$ of $i$-th layer and the output $h_{i-1} \in R^{b \times l \times d_{i-1}}$ of the previous layer, we assume that these values of all weights in this weight matrix is the big but same, where $b$ is batch size, $l$ is the maximum length of the sequences of user-item interactions. When we use this layer to capture the dependency of the input sequence and output $h_{i} \in R^{b \times l \times d_{i}}$, the each value of the dimension $h_{i}^{j,k} \in R^{d_{i}}$ of the output is also the same and meaningless, but the $H(W_i)$ is big using L1-norm or L2-norm criterion, where $j \in \{1,…,b\}$, $k \in \{1,…,l\}$ and $H$ indicates the degree of the information of weight matrix.
To address the limitation mentioned above, we introduce an entropy-based criterion to measure the variation of  weights in each layer. Entropy is often used to evaluate the degree of chaos (information) in a system~\cite{fletcher2018inference, reeves2017additivity}. 
 Inspired by~\cite{strong1998entropy, shwartz2017opening, meng2020filter},
% Obviously, the above example $H(W_i)=0$ that in line with our inference. Based on the above analyses, we chose entropy as the criterion for measure the importance of each layer information of model. 
% \deleted{There are several methods \cite{gabrie2018entropy, kraskov2004estimating} to implement the entropy estimation, which can be classified as parametric methods non-parametric methods. Parametric methods \cite{gabrie2018entropy} need large scale samples to train DNN-based model, which is not unacceptable for our task due to the sample (i.e. weight matrix) is lacking and the unknowable of optimal parameters of model. Non-parametric methods including binning-based methods and k-nearest neighbor (KNN) methods. Among them, k-nearest neighbour distances \cite{kraskov2004estimating} cannot efficiently obtain the entropy result of large-scale data and completely affects the training time of the model. To the contrary, binning based methods \cite{strong1998entropy, shwartz2017opening} can quickly calculate entropy even for high-dimensional data and efficiently evaluate how many weights matter (i.e. information) in a layer. Although it may lead to marginally systematic bias, which we use an adaptive coefficient to ease this problem.}
% When calculate the entropy of each layer, following these strategies
we transform the weight matrix into a vector and discretize the vector into $m$ bins.
Then we can calculate the probabilities of each bin.
% calculate the probabilities of each bin by discretizing the vector into $m$ bins.
To be specific, we first sort the weights in the vector based on their actual values and divide the vector into $m$ bins with equal numeric intervals ($\frac{\max - \min}{m}$ where $\max$ and $\min$ represent the maximum and minimum values of the weight matrix, respectively). The probability of the $j$-th bin is:
\begin{equation}
p_{j}=\frac{n_{j}}{N}
\end{equation}
where $N$ and $n_j$ are the parameter sizes of the weight vector and the $j$-th bin, respectively. Then, we calculate the entropy (information) of the weight matrix $W_i$ as follows:
\begin{equation}
\mathrm{H}(\mathrm{W}_{i})=-\sum_{k=1}^{m} p_{k} \log p_{k}.
\label{entropy}
\end{equation}
A smaller score of $\mathrm{H}(\mathrm{W}_{i})$ means the layer in this model has less variation (information).

\begin{algorithm}[t]

    \caption{Layer-wise Cooperation}
    \label{algorithm}
    % \Indm\Indmm
    % \KwIn{two models ${M}_1, {M}_2$, where each model has $N$ layers with weights ${W}_1^k, {W}_2^k$, $k = 1, 2$; two different shuffles $\mathcal{D}_1, \mathcal{D}_2$ on training data;
    % two different learning rates, $\gamma_1, \gamma_2$; 
    % the hyper-parameter $\alpha$;}\\
      \KwIn{${M}_1, {M}_2$ with weights ${W}_1^k, {W}_2^k, \dots, {W}_N^k$, $k = 1, 2$; $\mathcal{D}_1, \mathcal{D}_2$; $\eta_1, \eta_2$; 
    the hyper-parameter $\alpha$;\\}
    $t = 0.$ \\
    \SetAlgoNoLine
    \KwOut{ \\
        \Indp\Indpp
        $t = t + 1$ \\
        \For{k = 1, 2}{
            Compute the gradient and Update $M_k$ with $\eta_k$ and $\mathcal{D}_k$; \\
            \tcc{Update parameters at each batch for all models parallelly or serially.}
        }
        \For{k = 1, 2}{
            \For{i = 1, 2, \dots, N}{
                Calculate $H(W_i^k)$ and $H(W_i^{k+1})$ by Eq.~(\ref{entropy}) \;
                Calculate $\mu_i^k$ with $\alpha$
                % with $H(W_i^k)$, $H(W_i^{k+1})$ and $\alpha$
                by Eq.~(\ref{coefficient}) \;
                % $\widehat{W}_i^k = {\mu_i^k} * W_i^k + (1-{\mu_i^k})*W_i^{k+1}$ 
                 Get $\widehat{W}_i^k$
                by Eq.~(\ref{linearcombine}) ; 
                \tcc{Replace $ W_i^k$ with $\widehat{W}_i^k$ in $M_k$. }
                }
        }
    % \Repeat{}{}
    }
    \KwResult{convergence} 
\end{algorithm}

\subsection{PCRec Framework}

\begin{figure}[t]
  \includegraphics[width=0.43\textwidth]{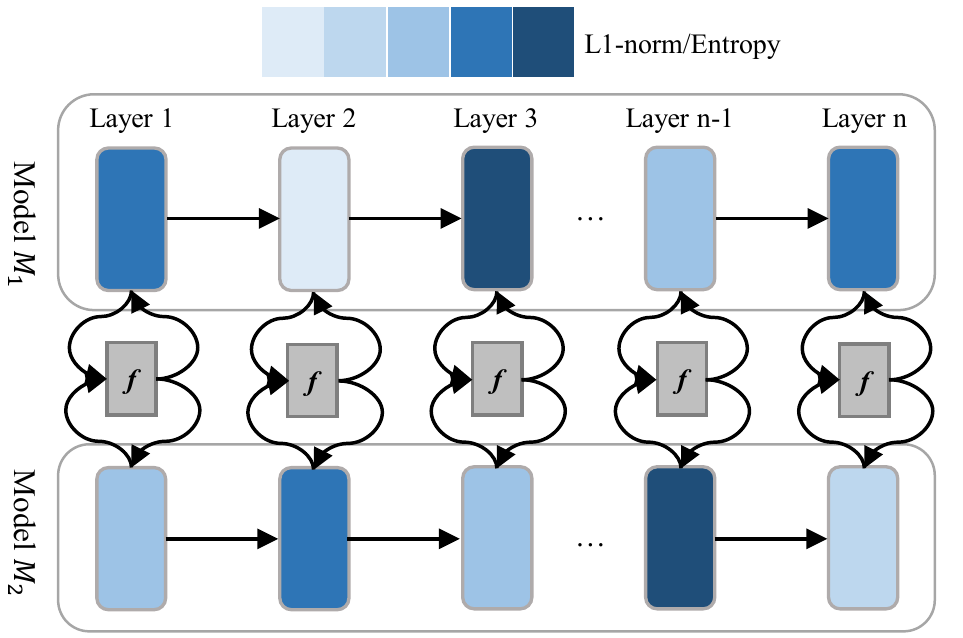}
  \caption{PCRec framework with LW cooperation.
 $f$ represents linear combination based on layer information, i.e., $H(W_i^1)$ and $H(W_i^2)$.
%  \added{Through LW-cooperation, we reactivate the invalid weights of all layers to enhance the representation learning of item(user).}
%  \deleted{$t$ denotes the size of feature (e.g., user-item interactions) space, $y$ is very flexible, which can represent a prediction score or an output distribution or an item sequence, as illustrated in section~\ref{Instantiationsection}}. 
%   For example, taking the sequential recommender model NextItNet as an example, given the input interaction sequence $\{x_{1}, x_{2},$ $ ..., x_{t}\}$,  the prediction of its output should be  $\{x_{2}, x_{3},$ $ ..., x_{t+1}\}$. 
%   As an example, each single network depicted here is a self-supervised autoregressive model, e.g., NextItNet.
}
  \label{framework}
\end{figure}

 We illustrate the proposed PCRec framework in Figure~\ref{framework}.
%  \added{, which covers two rules including individual weight and all weights in a neural network layer as the measure unit respectively}.
Assume that both models have $N$ layers. We denote $W_{i}^{1}$ and $W_{i}^{2}$
% \in  \mathbb{R}^{d_{i-1} \times d_{i}} 
as the weight matrices of the $i$-th layer of the two models.
% $H(W_{i}^{1})$ and $H(W_{i}^{2})$ as the information of $W_{i}^{1}$ and $W_{i}^{2}$, respectively. 
Our core idea is to use the corresponding weight information of the two networks, and generate more expressive weights $\widehat{W}_{i}$ as
\begin{equation}
\widehat{W}_{i}=f(W_{i}^{1}, W_{i}^{2}). 
\end{equation}
% The importance of the $W_{i}^{1}$ and $W_{i}^{2}$ is relative.
The weight $W_{i}^{1}$ and $W_{i}^{2}$ are significantly different since they are optimized with different hyper-parameters (mentioned later) and initialization. That is, the unimportant weights of a layer may correspond to the important weights of the same layer in his peer, and vice versa. Before describing the layer-wise (LW) cooperation mechanism, we first show a more intuitive parameter-wise (PW) method by exploiting redundancy pruning.

\subsubsection{PW Cooperation}
% The core idea of pruning \cite{han2015deep} is to search invalid weights and remove them. We build on top of that approach. 
The process is shown in Figure~\ref{individualweight}, we first define a positive threshold $\gamma$ and then identify unimportant parameters if their absolute values are smaller than  $\gamma$. To realize information transfer from its peer model, we  simply replace 
 these unimportant parameters with parameters in its peer model of the same layer and index position. To realize this, we define a binary mask matrix
%  where the index are defined 0-1 matrix $
 $I_i^k \in  \mathbb{R}^{d_{i-1} \times d_{i}}$ which has the same shape with $W_{i}^{k}$ to indicate the indices of these invalid weights in $W_{i}^{k}$
%  \added{to indicate the indexes of these invalid weights in $W_{i}^{k}$}. 
%  Values in   $I_i^k$ are set to 1 if their absolute value are larger than $\gamma$, and 0 otherwise.
 This process is symmetrical for the two peer models.  Correspondingly, we can formulate the PW process as follows.
 \begin{equation}
 \label{pw-newweight}
\begin{array}{c}
\widehat{W}_{i}^{k}=W_{i}^{k} * (1-I_{i}^{k})+W_{i}^{k \% 2+1} *I_{i}^{k}, \quad k=1,2.
\end{array}
\end{equation}
where \% is the modulo operation and each element of $I_{i}^{k}$ is:
%  \begin{equation}
% \begin{array}{c}
% I_{i}^{k}=W_{i}^{k}<\text { threshold }
% \end{array}
% \end{equation}
%  \begin{equation}
% I_{i, m, n}^{k}=\left\{\begin{array}{ll}
% 1 & \text { if } W_{i, m, n}^{k} \geq \text { threshold } \\
% 0 & \text { if } W_{i, m, n}^{k}<\text { threshold }
% \end{array}\right.
% \end{equation}

\begin{equation}
\label{index}
I_{i, m, n}^{k}=\left\{\begin{array}{ll}
0 & \text { if } W_{i, m, n}^{k} \geq \gamma \\
1 & \text { if } W_{i, m, n}^{k}<\gamma
\end{array}\right. 0 \leq m<d_{i-1} \text { and } 0 \leq n<d_{i}
\end{equation}
The learning process of PCRec with the PW cooperation is illustrated in Algorithm~\ref{pw_alg}. While this PW cooperation is intuitively simple, it has some shortcomings as mentioned below. 
 
% we capture these important weights with bigger value by L1-norm and capture invalid indexes of other invalid weights from the weight matrix $W_{i}^{1}$, where the invalid weights are below a threshold and vice versa. Then, we combine these important weights with the transplanted information from the peer model as the new and reactivated weight matrix $\widehat{W}_{i}^{1}$, where the transplanted information are captured by utilizing the invalid indexes in the peer weight matrix $W_{i}^{2}$. Noted that the rule is symmetrical for two peer model. We formulate the rule as:

\subsubsection{LW Cooperation}
% \deleted{Our core idea is to use the corresponding weight information of the two networks, and generate more expressive weights $\widehat{W}_{i}$ as
% \begin{equation}
% \widehat{W}_{i}=f(W_{i}^{1}, W_{i}^{2}). 
% \end{equation}
% % The importance of the $W_{i}^{1}$ and $W_{i}^{2}$ is relative.
% The weight distributions of $W_{i}^{1}$ and $W_{i}^{2}$ could be different since they are optimized with different hyper-parameters (mentioned later). That is, the unimportant weights of a layer may correspond to the important weights of another layer, and vice versa. }
% \deleted{Noted that this situation does not hinder the two layers can capture the same representation for instance.} 
Using individual weight as the measure unit only focuses on the importance of the weight itself, which unfortunately ignores the layer consistency and may thus hurt the model expressivity and performance. We argue that using the entire layer as the measure unit can enable all weights at the same layer to contribute synergistically to the transformation of the layer.
% A problem of individual weight measure unit is that it neglects the connectivity of the weights,
% This may hurt the distribution of weights and even hurt the model performance. 
Thus, we propose a layer-wise transplanting method by defining $f$ as a linear combination function:
\begin{equation}
\begin{array}{l}
\widehat{W}_{i}^{k}={\mu_i^k} W_{i}^{k}+(1-{\mu_i^k}) W_{i}^{{k\%2}+1}, \quad k=1,2.
\end{array}
\label{linearcombine}
\end{equation}
where $0 \leq \mu_i^k \leq 1$ is the coefficient. 
% and $\widehat{W}_{i}^{k}$ is the new weight matrix after information transplanting.
% Note that the new weight matrix was updated by using the two non-updated weight matrices. 
Particularly, we treat this coefficient $\mu_i^k$ as an adaptive parameter so as to promote cooperation and optimization automatically.
% \deleted{{\color{red}Note that regarding the bias terms or the normalization layers (i.e., gain and bias), we use $\mu_i^k$ which is calculated based on their corresponding feedforward or convolutional layers.}}
% to identify how much external information from other models is required, rather than needed to be tuned manually. 
Below, we give two instructions on designing a suitable adaptive parameter $\mu_i^k$:
%guideline of design  There are two emphases to consider  when deigning $\mu_i^k$:
\begin{itemize}
\item[(1)] We expect that layers with less information could get additional information from its peer model. Hence, we use the difference  $H(W_{i}^{1})-H(W_{i}^{2})$ to measure the relative importance of information in the two layers. When the difference is zero,  $\mu_i^k$ should be set to $0.5$, otherwise  $\widehat{W}_{i}^{k}$ should assign a large $\mu_i^k$ (i.e., $\mu_i^k > 0.5$) to the layer that has more information. Note that even $\mu_i^k=0.5$ could be also helpful since the same information does not mean all weights are identical according to Eq.~\eqref{entropy}. Consider an extreme situation where the distributions (e.g., normal distribution) of weight matrices are identical, but the magnitude of each weight (with the same position) is the opposite. In such a case, the information of each layer is the same, but the entropy of $\widehat{W}_{i}^{k}$ is enlarged by Eq.~(\ref{linearcombine}).
% Noted that the situation of the complementary phenomenon, 0.5 of coefficient is also hopeful to revise these invalid weights.
\item[(2)] Even the difference $H(W_{i}^{1})-H(W_{i}^{2})$ is large, we expect that $\widehat{W}_{i}^{k}$ contains part information of itself and is able to adaptively control the impact of  $H(W_{i}^{1})-H(W_{i}^{2})$.

% \deleted{We need to adaptively control of the coefficient based on the difference.}
% The different relative information with the same difference should be distinguished, which means the coefficient based on the same difference is not fixed. For example, when the different of  $H(W_{i}^1 )=0.9$ and $H(W_{i}^2)=0.4$ is 0.5, we may think the coefficient $\mu$ should be infinitely close to 1. But the different of $H(W_{i}^1 )=0.6$ and $H(W_{i}^2)=0.1$, the coefficient $\mu$ should be only close to 0.5.
\end{itemize}

To meet the above requirements, we design an adaptive $\mu_i^k$ which is wrapped by the sigmoid function:
\begin{equation}
\mu_i^k=\frac{1}{1+exp({-\alpha(H(W_{i}^{k})-H(W_{i}^{k\%2+1}))})} \quad k=1,2
\label{coefficient}
\end{equation}
where $\alpha$ is a hyper-parameter to control the degree of the information from the outside layer. It is worth noting that the calculation criterion of information for a layer as the measure unit can be L1-norm (i.e., Eq.~\eqref{indiweight}) or entropy (Eq.~\eqref{entropy}), which is different from the individual weight as the measure unit with L1-norm criterion. During training, we just need to perform this combination operation at each epoch. The new weight matrices ($\widehat{W}_{i}^1$ and $\widehat{W}_{i}^2$) should be the same for the two individual models due to the dual linear combination. 
In practice, we need to guarantee that each  model has diverse and sufficient information so as to complement each other. 
In this paper, we adopt two simple strategies for the two models to make each of them capture unique information, i.e.,
using different learning rates and sampling of the training data.
% which are using different learning rates and train data with different order to training the two models. We will examine the two strategies to valid our estimates in the experimental results~\ref{}. 

% \subsection{Extend to multiple networks}
% % \begin{figure}[t]
% %   \centering
% %   \includegraphics[width=0.45\textwidth]{reverse_prune.pdf}
% %   \caption{Multi-model cooperation framework, where we take four selfsame networks as an example.}
% %   \label{multiplemodels}
% % \end{figure}
% \deleted{The proposed two-model cooperation framework can be easily extended to multiple models, as shown in Figure~\ref{multiplemodels}. Given $K$ networks $M_{1}, M_{2}, \ldots, M_{K}$ ($K$>2).   The model $M_{i}(1 \leq \mathrm{i} \leq K-1)$ takes the next model $M_{i+1}$ as its partner, and correspondingly, the last model $M_{K}$ takes the first model $M_{1}$ as a partner so as to form a loop. That is, we use the parameters of $M_{i+1}$ as external information to complement $M_{i}$. Our multiple model cooperation framework is asymmetric. 
% % Equation~\ref{linearcombine} is a special case of PCRec with $K = 2$. 
% By this way, each recommender model can indirectly captures the external information from the rest models with the same architecture. }
% \begin{equation}
% \begin{array}{l}
% \widehat{W}_{i}^{k}={\mu_i^k} W_{i}^{k}+(1-{\mu_i^k}) W_{i}^{{k\%K}+1} \quad k=1,\dots,K
% \end{array}
% \label{multi-linearcombine}
% \end{equation}

\begin{table}[t]
\small
	\caption{\small Statistic of the evaluated datasets. "M" and "K" is short for million and kilo, "t" is the length of interaction sequences.}
	\setlength{\tabcolsep}{4mm}
	%\scalebox{0.9}{
		\begin{tabular}{lccccc}
			\hline
			Dataset        & \#users & \#items & \#actions & t \\ 
% 			\hline
            \toprule[0.01mm]
            \bottomrule[0.01mm]
			Retailrocket   & 47K  & 61K  & 0.4M        & 10        \\
			QQBrowser      & 978K  & 70K  & 25M        & 50  \\
			ML-20M         & 138K  &18K  & 20M     & 100   \\
 
			\hline
		\end{tabular}
	%}
\label{datadescribe}
\vspace{-0.2in}
\end{table}

\subsection{Model Optimization}
%Although PCRec uses multiple models to capture optimal performance, which is similar to ensemble learning and knowledge distillation (KD). There are fundamental differences between them: (1) PCRec utilizes the information of weights of model to complement each other, instead of the outputs or intermediate representations of ensemble learning or KD. (2) PCRec only uses one model to predict during testing, which like KD, but different from ensemble learning. (3) PCRec is designed for improving the model performance as ensemble learning, and KD is for model compression. Since the purpose of ensemble learning is the same as our PCRec, we will further study in the experimental results~\ref{}.

% \replaced{Give several comments or detailed descriptions on the procedures for above Algorithm \ref{algorithm}}{}

% \DecMargin{1.5em}
PCRec can be optimized in two modes, namely, parallel and serial training. In terms of the parallel mode, the  two individual networks of PCRec are essentially trained independently, but each batch of them is trained concurrently.
% , which equal with the cost regarding of time of training one model.
The information of each identical recommender model can be transferred by using the saved checkpoint. As a result, parallel optimization requires more memory and computations, but saves substantial training time. For clarity, if we assume the time and space complexity of each model are the scalars $TC$ and $SC$, the time and space complexity of PCRec in parallel training mode are $TC$ and $2SC$. On the other hand, we can perform serial optimization for each individual network by sequentially training them per batch. As such, compared with the parallel mode, the serial optimization inevitably sacrifices training time but consumes no extra memory and computation. The time and space complexity of PCRec in the serial training model are roughly $2TC$ and $SC$. Algorithm~\ref{algorithm} illustrates the peer cooperation process. In summary, we maintain two networks with an identical structure but different learning rates and sampling orders. When a training epoch is finished, we calculate the information $H(W_i^k)$ of each layer of the two models and perform cooperation. Note that the parameters of the bias and normalization terms of the same layer share the same $\mu_i^k$ calculated based on $W_i^k$.
% Regarding the training loss, we apply the cross-entropy loss~\cite{yuan2019simple} for all individual models in this paper.
After training, PCRec needs only one 
peer model for inference, and thus, is as efficient as the original individual recommender model. This property is distinct from the traditional ensemble methods that have to rely on the decisions of multiple `weak' learners during inference.

\section{EXPERIMENTAL SETUP}
\label{expsetup}
We describe the experimental setup in this section, including datasets, baselines,  implementation details and evaluation metrics. 

\subsection{Datasets}
\begin{itemize}
	\item{\textbf{ML-20M\footnote{https://grouplens.org/datasets/movielens/20m/}}}: 
	This is a well-known benchmark dataset widely used for both traditional and sequential recommendation tasks~\cite{kang2018self,sun2019bert4rec,sun2020generic}. It contains around 20 million user-item interactions with 27,000 movies and 138,000 users.  Following the common practice in \cite{yuan2019simple, yuan2020future, kang2018self}, we assume that an observed feedback is available if an explicit rate is assigned to this item.
	We perform basic pre-processing to filter out the interactions with less than 5 users and users with less than 5 items to alleviate the effect of cold users and items. Then, we use timestamps to determine the order of interactions. Following~\cite{kang2018self, sun2019bert4rec}, we adopt the leave one out evaluation scheme. For each user, we hold out the last item of the interaction sequence as the test data, treat the item just before the last as the validation set, and utilize the remaining items for training. For the sequential recommendation task, we construct user's interaction sequences by using his recent $t$ interactions by the chronological order. For sequences shorter than t, we simple pad them with zero at the beginning of the sequence following~\cite{yuan2019simple}, while for sequences longer than t, we split them into several sub-sequences with length $t$ in the training set. In this paper, we set $t$ to 100 on this dataset. 
% 	The recommender model aims to predict the next interaction by using  his/her previous $t-1$ behaviors as features.
\begin{table}[t]
\setlength{\tabcolsep}{0.35mm} % 
\small
\caption{Hyper-parameter details.}.
\begin{tabular}{l|ccccc|ccccc|ccccc}
\hline
\textbf{Data}  & \multicolumn{5}{c|}{\textbf{Reatilrocket}}                       & \multicolumn{5}{c|}{\textbf{ML-20M}}                              & \multicolumn{5}{c}{\textbf{QQBrowser}}                          \\ \hline
\textbf{Model} & \textbf{$b$} & \textbf{$d$} & \textbf{$\eta$} & \textbf{$L_2$} & \textbf{$p$} & \textbf{$b$} & \textbf{$d$} & \textbf{$\eta$} & \textbf{$L_2$} & \textbf{$p$} & \textbf{$b$} & \textbf{$d$} & \textbf{$\eta$} & \textbf{$L_2$} & \textbf{$p$} \\ 
% \hline
\toprule[0.01mm]
\bottomrule[0.01mm]
SASRec         & 128        & 64         & 1e-3       & -           & 0.3        & 128        & 256        & 1e-3       & -           & 0          & 128        & 256        & 1e-3       & -           & 0.5        \\
DNN     & 128        & 64         & 1e-4      & 1e-5        & -          & 128        & 256        & 1e-4      & 1e-6        & -          & 128        & 256        & 1e-4      & 1e-5        & -          \\
BPR            & 2048       & 256        & 1e-3       & 1e-4        & -          & 2048       & 256        & 1e-3       & 0           & -          & 2048       & 256        & 1e-3       & 1e-4        & -          \\ \hline
\end{tabular}
\label{hyperpara}
\end{table}

\begin{table*}[t]
\setlength{\tabcolsep}{0.58mm} % 
\caption{Overall performance of all models. PCRec with two SASRec, DNN and BPR is referred to {PC-SAS}, {PC-DNN} and {PC-BPR}, respectively. Here, we present the results of PCRec with LW-cooperation and entropy-based information criterion because of its best performance.
% and PC-SAS and PC-DNN only adopt LW-cooperation in the embedding layer
We set $\alpha$ of PC-SAS to 30, 30, 30, $\alpha$ of PC-DNN to 40, 40, 10, and $\alpha$ of PC-BPR to 20, 20, 20, on Retailrcoket, ML-20M, QQbrowser, respectively. Improvements over baselines are statistically significant with p < 0.01.}
\begin{tabular}{c|c|c|c|c|c|c|c|c|c|c|c|c}
\hline
Data           & \multicolumn{4}{c|}{\textbf{Retailrocket}}                                     & \multicolumn{4}{c|}{\textbf{ML-20M}}                                  & \multicolumn{4}{c}{\textbf{QQBrowser}}                               \\ \hline
\textbf{Model} & \textbf{MRR@5}  & \textbf{MRR@20} & \textbf{HIT@5}  & \textbf{HIT@20} & \textbf{MRR@5}  & \textbf{MRR@20} & \textbf{HIT@5}  & \textbf{HIT@20} & \textbf{MRR@5}  & \textbf{MRR@20} & \textbf{HIT@5}  & \textbf{HIT@20} \\ 
% \hline
\toprule[0.01mm]
\bottomrule[0.01mm]
BPR            & 0.0599          & 0.0712          & 0.1091          & 0.2321          & 0.0250          & 0.0332          & 0.0483          & 0.1412          & 0.0184          & 0.0239          & 0.0352          & 0.0953          \\
PC-BPR         & 0.0650          & 0.0770          & 0.1179          & 0.2420          & 0.0274          & 0.0360          & 0.0531          & 0.1530          & \textbf{0.0208} & \textbf{0.0264} & \textbf{0.0392} & 0.1041          \\ \hline
DNN     & 0.1648          & 0.1704          & 0.2087          & 0.2628          & 0.0334          & 0.0415          & 0.0628          & 0.1484          & 0.0123          & 0.0165          & 0.0246          & 0.0709          \\
PC-DNN         & 0.1706          & 0.1777          & 0.2198          & 0.2900          & 0.0364          & 0.0445          & 0.0671          & 0.1531          & 0.0133          & 0.0178          & 0.0265          & 0.0756          \\ \hline
SASRec         & 0.2022          & 0.2192          & 0.3169          & 0.4830          & 0.1292          & 0.1443          & 0.2049          & 0.3593          & 0.0185          & 0.0250          & 0.0371          & 0.1055          \\
PC-SAS         & \textbf{0.2091} & \textbf{0.2261} & \textbf{0.3262} & \textbf{0.4921} & \textbf{0.1360} & \textbf{0.1513} & \textbf{0.2137} & \textbf{0.3710} & 0.0195          & \textbf{0.0264}          & \textbf{0.0392}          & \textbf{0.1111} \\ \hline
\end{tabular}
\label{overallresult}
\end{table*}

	\item{\textbf{QQBrowser\footnote{https://browser.qq.com/}}}: It is an industrial dataset which was collected from the QQBrowser platform of Tencent. The items in QQBrowser include news, videos and ads. It consists of more than 70,000 items and almost 1 million users. We perform a similar pre-processing as above and set $t$ to 50. We will open source this dataset later for reproducibility.
	
	\item{\textbf{Retailrocket\footnote{https://www.kaggle.com/retailrocket/ecommerce-dataset}}}: It is a public dataset collected from a real-world ecommerce website,  consisting  user shopping behaviors in 4.5 months. It contains 235,061 items and 1.4 million users. Similarly, we set $t$ to 10 to investigate recommendation performance for short-range interaction sequences.
\end{itemize}
Table~\ref{datadescribe} summarizes the statistics of evaluated datasets after the basic pre-processing.

% \begin{table}[t]
% \small
% 	\caption{\small Statistic of the evaluated datasets. "M" and "K" is short for million and kilo, "t" is the length of interaction sequences.}
% 	\setlength{\tabcolsep}{4mm}
% 	%\scalebox{0.9}{
% 		\begin{tabular}{lccccc}
% 			\hline
% 			Dataset        & \#items & \#actions & \#sequences & t \\ \hline
% 			Weishi         & 66K   & 10M     & 1048575   & 10        \\
% 			ML30           & 27K   & 20M     & 711771   & 30   \\
% 			QQBrowser      & 70K   & 26M     & 979777   & 50  \\
% 			ML70           & 27K   & 20M     & 349501    & 70   \\
% 			\hline
% 		\end{tabular}
% 	%}
% \label{datadescribe}
% \vspace{-0.1in}
% \end{table}

\subsection{Baseline model}
We evaluate the PCRec framework by using three popular recommender models, namely, SASRec~\cite{kang2018self},  YouTube DNN~\cite{covington2016deep} (DNN for short) and BPR~\cite{rendle2012bpr}. For SASRec, we use its official code online\footnote{\url{https://github.com/kang205/SASRec}}, while for BPR and YouTubeDNN, we implement it by strictly following the original paper.
% \deleted{Besides, we use two powerful sequential recommendation system, GRU4Rec \cite{hidasi2015session} and SASRec \cite{kang2018self}, and non-sequential recommendation system FM \cite{rendle2010factorization} as baselines.} 
 It is worth noting that compared with SASRec, DNN and BPR are unable to capture user sequential patterns. This  is because DNN model user's previous interactions as common features, while BPR with matrix factorization as the scoring function is a typical collaborative filtering baseline. 
We want to emphasize that the purpose of our study is not to propose a state-of-the-art model beating existing baselines. The purpose is rather to introduce a new learning paradigm that could effectively leverage the parameter redundancy issues in large and deep recommender models  so as to achieve some additional improvement in accuracy.

\subsection{Implementation details}
We train all models using the Adam optimizer on GPU. For common hyper-parameters, we consider the hidden dimension size (denoted by $d$) from \{16, 32, 64, 128, 256\} and the learning rate (denoted by $\eta$) from \{0.0001, 0.00025, 0.0005, 0.00075, 0.001, 0.005\}, the $L_2$ regularization coefficients from \{0.01, 0.001, 0.0005, 0.0001, 0.00005 0.00001\}, and dropout rate (denoted by $p$) from \{0, 0.1, 0.2, \dots, 0.9\} by grid search in the performance of the validation set. Specifically, we set the $d$ 256 for SASRec (except on Retailrocket), DNN (except on Retailrocket) and BPR. On Retailrocket, $d$ of SASRec and DNN is set to 64 to prevent overfitting. 
We use $\eta$ 1e-3 for SASRec and BPR, and 1e-4 for DNN on all datasets. In addition, we set batch size (denoted by $b$) to 128 for SASRec and DNN, and 2048 for BPR because of its enormous triple samples. As for model-specific hyper-parameters, we use two self-attention blocks (denoted by $l$) with one head for SASRec according to the original paper. Regarding DNN, we use one hidden layer on all datasets since using more layers does not lead to any improved results. 
Our PCRec uses exactly the same hyper-parameters (except $\eta$) as these individual base models. For $\eta$, one peer in PCRec uses exactly the same one with its base model, while the other peer uses a sub-optimal $\eta$. 
The model-specific hyper-parameter of PCRec  $\alpha$ is studied in the ablation study part. Without special mention, we report our results with the optimal $\alpha$. Detailed hyper-parameters are reported in Table~\ref{hyperpara}.

\subsection{Evaluation Metrics}
We follow previous works~\cite{yuan2019simple, yuan2020future, sun2020generic, kang2018self} by comparing the top-N metrics, namely, MRR@N(Mean Reciprocal Rank), HR@N(Hit Ratio) and NDCG@N(Normalized Discounted Cumulative Gain). To save space, we omit the formulas of these metrics. 
% Note that since we only have one test item for each user, Hit@N is equivalent to Recall@N, and is proportional to Precision@N. MRR is equivalent to Mean Average Precision(MAP). 
% For all these metrics, the higher the value, the better the performance. 
N is set to 5 and 20 in this paper. 
% \deleted{Similar to \cite{yuan2019simple, sun2020generic}, we evaluate the prediction accuracy of the last user-item interaction in the user sequence of the testing set.}
\vspace{-0.1in}

\begin{figure*}[t]
    \centering
    \begin{subfigure}[t]{0.24\linewidth}
    \centering
            \includegraphics[width=1.67in]{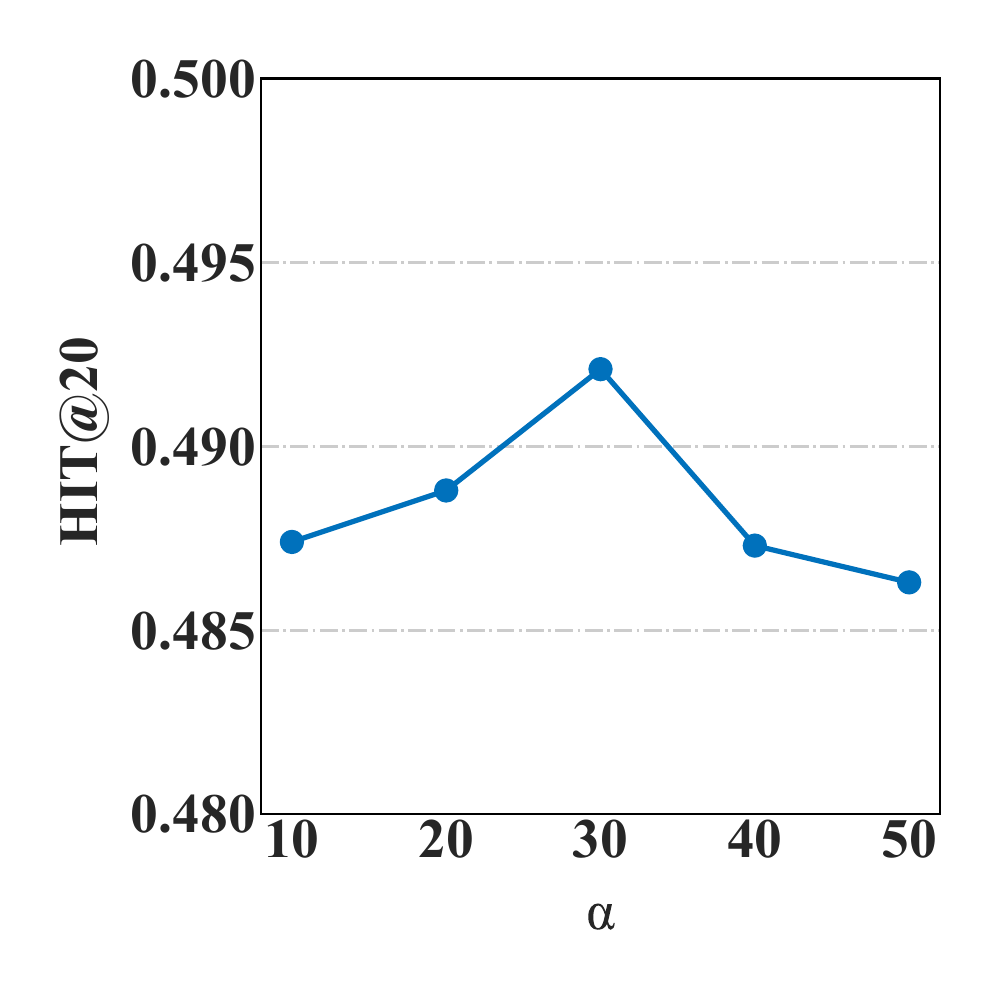}
            \subcaption{PC-SAS on Retailrocket}
            \label{aw}
    \end{subfigure}%
    \begin{subfigure}[t]{0.24\linewidth}
    \centering
            \includegraphics[width=1.67in]{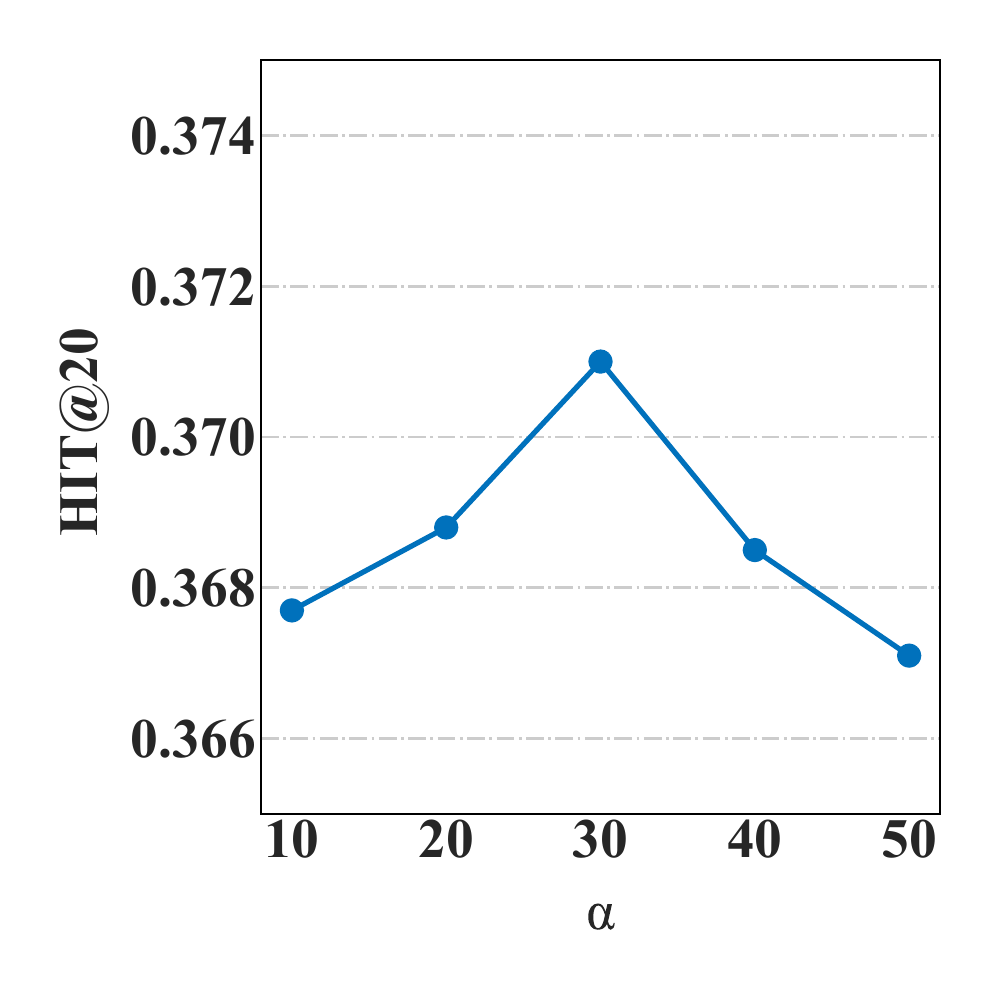}
            \subcaption{PC-SAS on ML-20M}
            \label{am}
    \end{subfigure}
    \begin{subfigure}[t]{0.24\linewidth}
    \centering
            \includegraphics[width=1.67in]{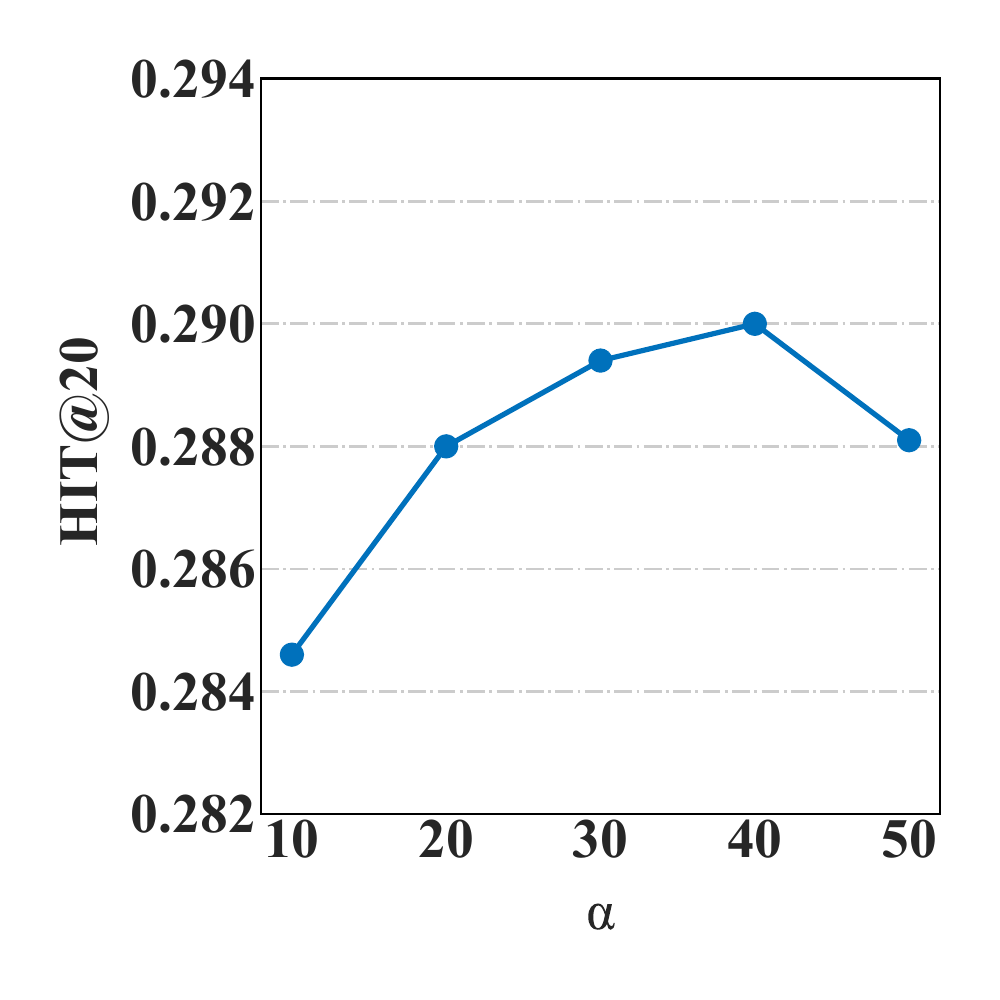}
            \subcaption{PC-DNN on Retailrocket}
            \label{anw}
    \end{subfigure}
    \begin{subfigure}[t]{0.24\linewidth}
    \centering
            \includegraphics[width=1.67in]{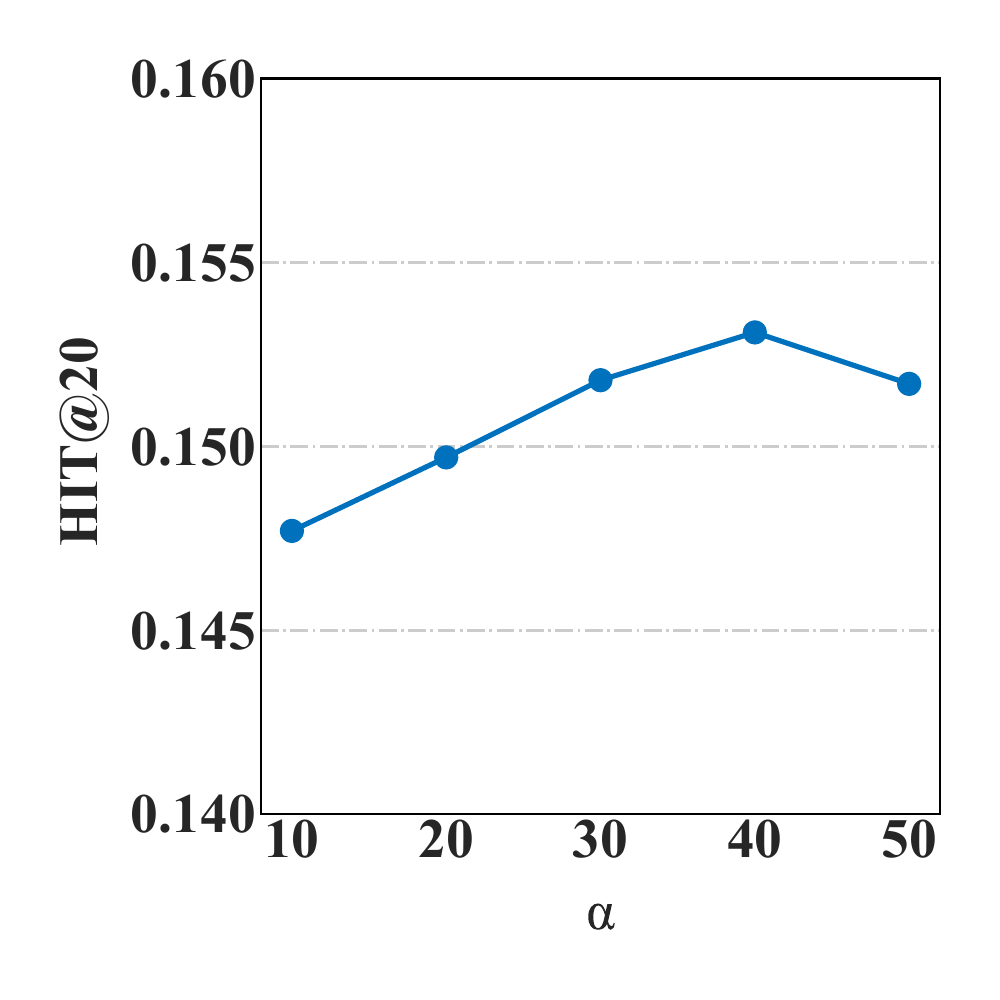}
            \subcaption{PC-DNN on ML-20M}
            \label{anm}
    \end{subfigure}
    % \vspace{-0.15in}
    \caption{The impact of $\alpha$ for PCRec.
    % : (a) the prediction accuracy (mrr@5) with respect to the $\alpha$ of SeCo-Next on Weishi; (b) the prediction accuracy (mrr@5) with respect to the $\alpha$ of SeCo-Next on ML30; (c) the prediction accuracy (mrr@5) with respect to the $\alpha$ of SeCo-NFM on Weishi; (d) the prediction accuracy (mrr@5) with respect to the $\alpha$ of SeCo-NFM on ML30;
    }
    \label{hyper-alpha}

\end{figure*}

\section{EXPERIMENTAL RESULTS}
\label{experimentsetup}
% The key contribution of PCRec is to improve the performance for recommender models based on deep neural networks.
In this section, we would answer the following research questions:
\begin{itemize}
\item \textbf{RQ1}: Does PCRec improve the performance of these typical neural networks, such as SASRec, YouTubeDNN and BPR?

\item \textbf{RQ2}: What is the performance of these variants of PCRec, which include PCRec with PW and LW cooperation, PCRec with L1-norm and entropy criteria.
% using individual weight and layer as measure unit with L1-norm criterion and the layer as measure unit with entropy criterion, respectively?

\item \textbf{RQ3}: What is the impact of the collaboration of different components in PCRec, such as, the embedding layer, softmax layer and hidden layers? 

\item \textbf{RQ4}: What is the impact of $\alpha$ for PCRec? Are the different learning rates and training data orders necessary?
% \item[(5)] \textbf{RQ4}: What is the influence of the different embedding size for PCRec?

\item \textbf{RQ5}: Does PCRec really enhance these unimportant weights of the original model? % \item[(7)] \textbf{RQ6}: 

\end{itemize}

% \vspace{-0.15in}

\subsection{Overall Evaluation (RQ1)}
We present the overall results in Table~\ref{overallresult}.
First, we observe that SASRec performs better than DNN and BPR with notable improvements. To our surprise, on Retailrocket and ML-20M, SASRec achieves several times improvements on all these top-N metrics. By examining the real dataset, we find that there indeed exist some short sequence fragments (formed with 2$\sim$4 videos) on the two datasets, which can be observed from the actions of many users. Unfortunately, DNN and BPR are unable to model such sequential patterns, and thus yield much worse results than the state-of-the-art sequential recommendation model SASRec. 

Second, as expected, PCRec, including PC-SAS, PC-DNN and PC-BPR, outperforms their individual base models (i.e., SASRec, DNN and BPR), demonstrating the effectiveness of peer collaboration. For example, compared with SASRec, PC-SAS achieves around 5\%  improvement in terms of MRR@5 on ML-20M; Compared with BPR, PC-BPR obtains up to 8\% improvement regarding MRR@5 on Retailrocket and ML-20M. In particular, PC-BPR outperforms BPR with around 11\% improvement regarding HIT@5 on QQBrowser. Notable improvements can also be observed by comparing PC-DNN to DNN on all datasets and all metrics.
In what follows, we would conduct ablation studies to verify the effectiveness of PCRec. To save space, we could only show partial results if the performance trends of them keep consistent.

\begin{table}[t]
\setlength{\tabcolsep}{2.2mm}
\caption{
The Comparison of PCRec variants. The standard ensemble learning method~\cite{rokach2010ensemble}  by averaging the prediction scores of two individual models is denoted by Ensemble-M2.
% PCRec can utilize layer or a single weight as the measure unit.
% PC-SAS with a single weight and a layer as measure unit is respectively denoted by PC-SAS-W and PC-SAS-L. \added{PC-SAS with guassian noise is denoted by PC-SAS-N and} PC-SAS-L with L1-norm and entropy criterion is denoted by PC-SAS-LL and PC-SAS-LE.  A similar expression also applies to PC-DNN and PC-BPR.
}
\begin{tabular}{c|cc|cc}
\hline
\textbf{Data} & \multicolumn{2}{c|}{\textbf{Retailrocket}} & \multicolumn{2}{c}{\textbf{ML-20M}} \\ \hline
\textbf{Model}         & \textbf{MRR@5}                & \textbf{HIT@5}               & \textbf{MRR@5}             & \textbf{HIT@5}            \\ 
%\hline
\toprule[0.1mm]
\bottomrule[0.1mm]
% BPR            & 0.0599               & 0.1091              & 0.0250            & 0.0483           \\
% PC-BPR-LE      & \textbf{0.0650}      & \textbf{0.1179}     & \textbf{0.0274}   & \textbf{0.0531}  \\
% PC-BPR-LN      & 0.0641               & 0.1171              & 0.0264            & 0.0517           \\
% PC-BPR-W       & 0.0638               & 0.1145              & 0.0234            & 0.0462           \\
% PC-BPR-N       & 0.0582               & 0.1068              & 0.0248            & 0.0478           \\ \hline
DNN     & 0.1648               & 0.2087              & 0.0334            & 0.0628           \\
PC-DNN-LE      & \textbf{0.1706}      & \textbf{0.2198}     & \textbf{0.0364}   & \textbf{0.0671}  \\
PC-DNN-LN      & 0.1686               & 0.2171              & 0.0355            & 0.0659           \\
PC-DNN-W       & 0.1665               & 0.2155              & 0.0351            & 0.0648           \\
PC-DNN-N       & 0.1639               & 0.2053              & 0.0325            & 0.0611           \\ Ensemble-M2    & 0.1693               & 0.2187              & 0.0355            & 0.0661           \\ \hline
SASRec         & 0.2022               & 0.3169              & 0.1292            & 0.2049           \\
PC-SAS-LE      & \textbf{0.2091}      & \textbf{0.3262}     & \textbf{0.1360}   & \textbf{0.2137}  \\
PC-SAS-LN      & 0.2078               & 0.3231              & 0.1343            & 0.2110           \\
PC-SAS-W       & 0.2071               & 0.3207              & 0.1338            & 0.2108           \\
PC-SAS-N       & 0.1994               & 0.3147              & 0.1283            & 0.2034           \\ 
Ensemble-M2    & 0.2076               & 0.3236              & 0.1357            & 0.2126           \\ \hline
\end{tabular}
\label{variantsresults}
% \vspace{-5pt}
\end{table}

\subsection{Comparison of PCRec Variants (RQ2)}
In Section~\ref{citeria}, we have proposed using a single weight and a layer as the measure unit in PCRec, We denoted them as PCRec-W and PCRec-L, respectively.
Further, in PCRec-L, we can adopt two criteria, L1-norm and entropy, to identify which layer of the two individual networks has less information, denoted as PCRec-LN and PCRec-LE, respectively. In addition,  we also evaluate a very simple method by reactivating the invalid weights using gaussian noise to increase the L1-norm, denoted as PCRec-N. 
We experimentally examine these methods and report results in Table ~\ref{variantsresults}. First, we find that  PCRec-N yields worse accuracy than the base model, which potentially indicates PCRec should use a useful information source, rather than random noise, for information transplanting. 
% clearly indicates that transplanting information from noise does not work at all. 
By contrast, PCRec-LE, PCRec-LN  always perform better than SASRec, DNN  on almost all datasets. This clearly verifies our main claim regarding the benefit of peer collaboration. 
% We also note that PCRec-W also shows 
% better results than the base model, but is obviously worse than PCRec-LN.
% We suspect that the parameter-wise information transplanting
%  may damage the layer consistency, and thus lead to relatively worse results than layer-wise cooperation.
Meanwhile,  PCRec-LN outperforms PCRec-W on most settings,  demonstrating the effectiveness of layer-wise cooperation;  
PCRec-LE outperforms PCRec-LN, demonstrating the effectiveness of entropy-based criterion, since it can more precisely identify how much information is required when performing information transplanting. 
% In practice, we strongly suggest developing PCRec by using layer-wise cooperation and entropy-based  information criterion.
 %  PCRec-LN and PCRec-W perform significantly worse than PC-SAS-LE with each layer as the measure unit and entropy criterion.
%  We argue that using layers as the unit is able to well maintain the layer consistency and using
%  entropy as criterion can more precisely identify how much information is required  when performing information transplanting.

On the other hand, we also compare the results that are produced by standard ensemble learning. It can be seen that the basic ensemble learning method (Ensemble-M2) is very effective and obviously surpasses these individual models. It even performs slightly better than PCRec-LN on the Retailrocket dataset when using DNN as the base model. However, our PCRec-LE in general can beat it, or at least they are competitive. Hence, we do not claim our PCRec is better than the standard ensemble learning method in this paper. But we emphasize that PCRec provides an alternative learning paradigm for getting information from an outside model, and more importantly, it is much more efficient than the standard ensemble learning during the inference phase, since it only requires one single model for prediction, rather than relying on predictions of two or more models. We further find that increasing the individual models for the ensemble learning, e.g., Ensemble-M3, does not yield better results.

% observe that 2.
% with two individual models could also achieve better performance than the single model (i.e., SASRec and DNN), demonstrating the effectiveness of model ensembling. In contrast with Ensemble-M2, SeCoRec achieves comparible results. But it is worth noting that SeCoRec is faster and requires less computation than the standard ensemble method during the inference phase, since it only uses one single model for prediction, rather than two models
% as in Ensemble-M2.

\begin{table}[t]
\setlength{\tabcolsep}{2.2mm}
\caption{
The impact of learning rates and sampling orders of training data for PCRec.
PC-SAS with different learning rates and sampling orders, the same learning rate and different sampling orders, different learning rates and the same sampling order, is denoted by PC-SAS-DD, PC-SAS-SD, PC-SAS-DS respectively. Similar expressions  apply to PC-DNN.}
\begin{tabular}{c|cc|cc}
\hline
\textbf{Data}  & \multicolumn{2}{c|}{\textbf{Retailrocket}} & \multicolumn{2}{c}{\textbf{ML-20M}} \\ \hline

\textbf{Model} & \textbf{MRR@5}       & \textbf{HIT@5}      & \textbf{MRR@5}    & \textbf{HIT@5}   \\ 
% \hline
\toprule[0.1mm]
\bottomrule[0.1mm]
DNN     & 0.1648               & 0.2087              & 0.0334            & 0.0628           \\
PC-DNN-DD      & \textbf{0.1706}      & \textbf{0.2198}     & \textbf{0.0364}   & \textbf{0.0671}  \\
PC-DNN-SD      & 0.1663               & 0.2119              & 0.0338            & 0.0628           \\
PC-DNN-DS      & 0.1681               & 0.2181              & 0.0352            & 0.0658           \\ \hline
SASRec         & 0.2022               & 0.3169              & 0.1292            & 0.2049           \\
PC-SAS-DD      & \textbf{0.2091}      & \textbf{0.3262}     & \textbf{0.1360}   & \textbf{0.2137}  \\
PC-SAS-SD      & 0.2059               & 0.3179              & 0.1317            & 0.2087           \\
PC-SAS-DS      & 0.2071               & 0.3218              & 0.1341            & 0.2113           \\ \hline
\end{tabular}
\label{rateandorder}
% \vspace{-0.2in}
\end{table}

\begin{figure*}[ht]
    \centering
    \begin{subfigure}[t]{0.24\linewidth}
    \centering
            \includegraphics[width=1.7in]{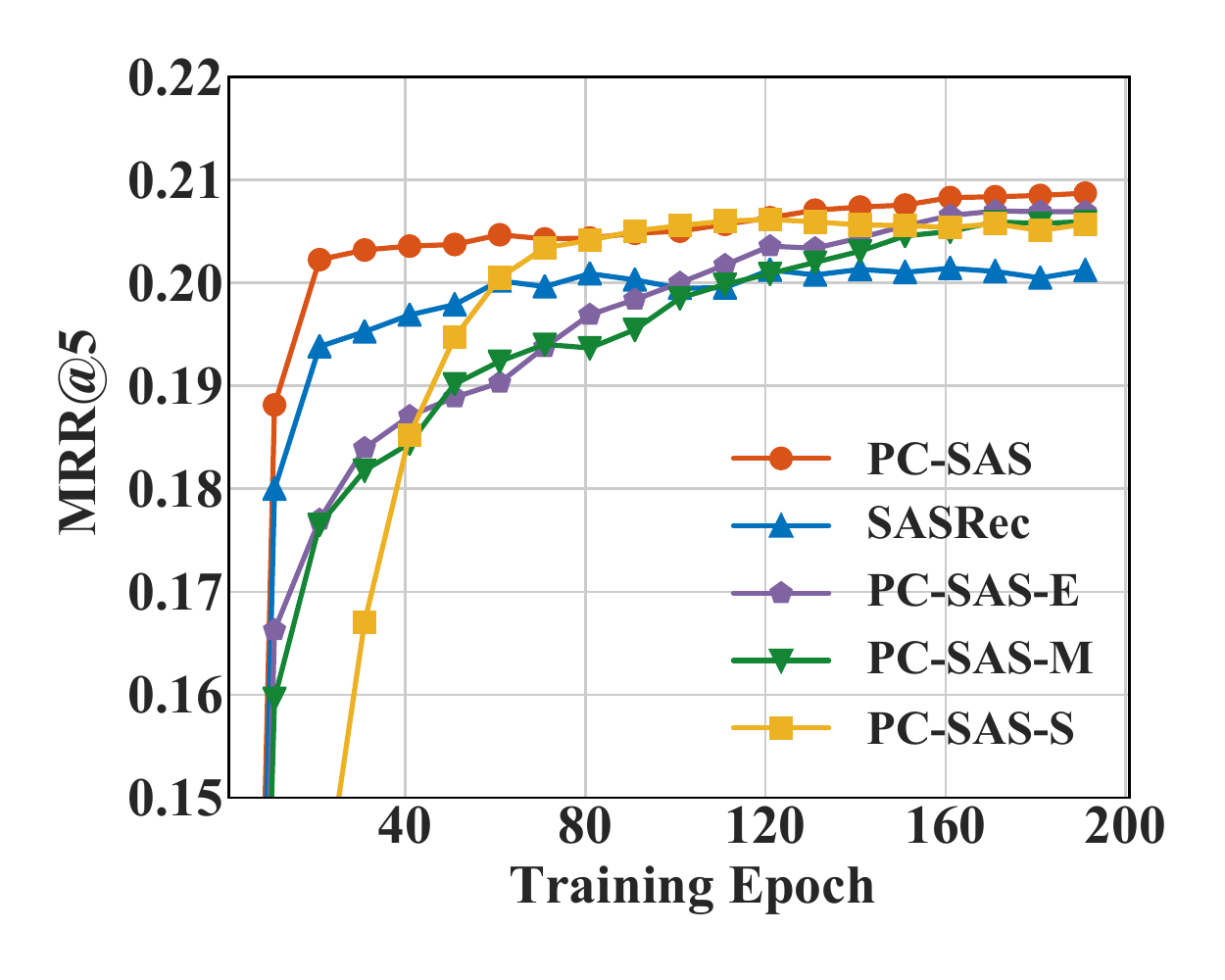}
            \subcaption{PC-SAS on Retailrocket}
    \end{subfigure}%
    \begin{subfigure}[t]{0.24\linewidth}
    \centering
            \includegraphics[width=1.7in]{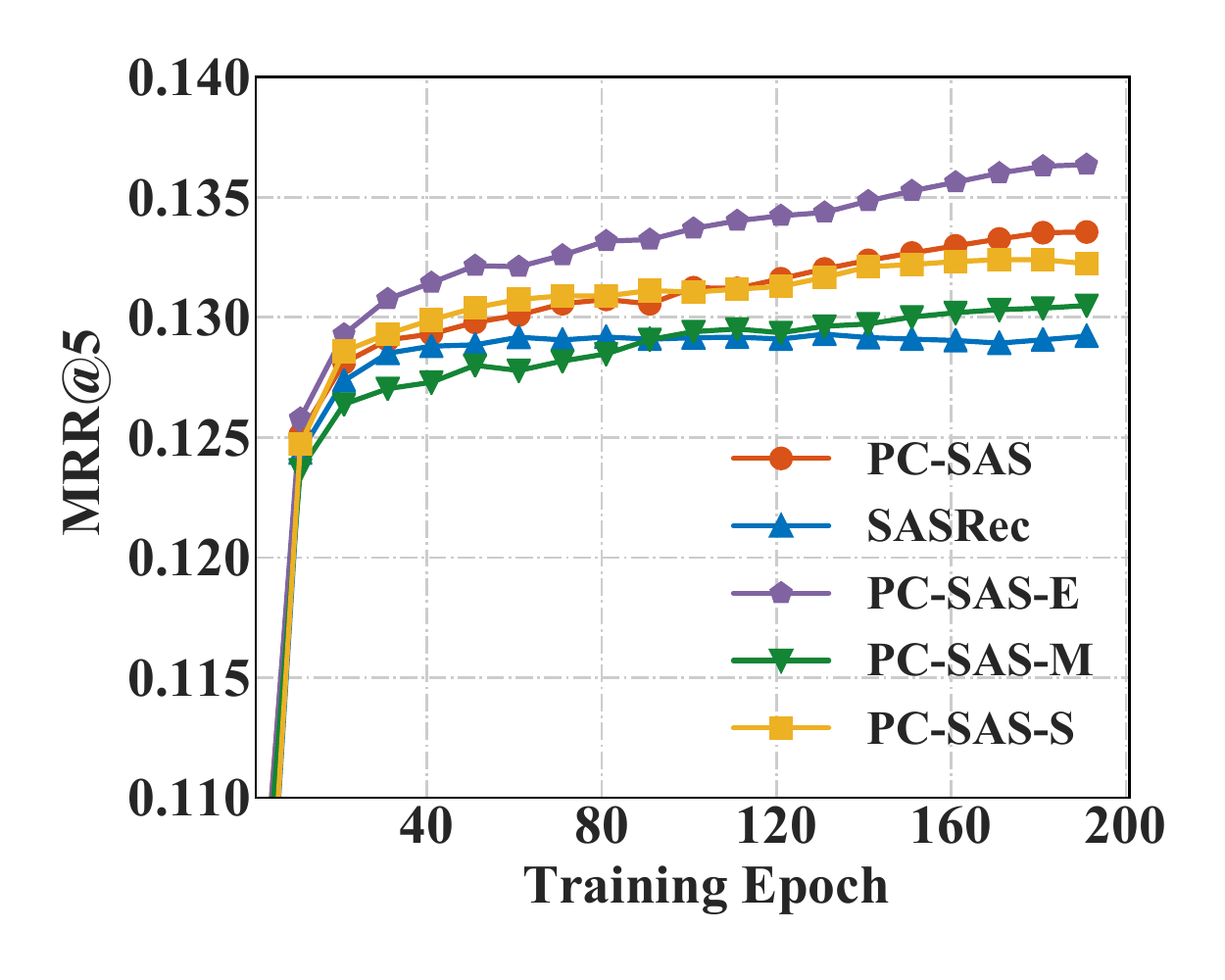}
            \subcaption{PC-SAS on ML-20M}
    \end{subfigure}
    \begin{subfigure}[t]{0.24\linewidth}
    \centering
            \includegraphics[width=1.7in]{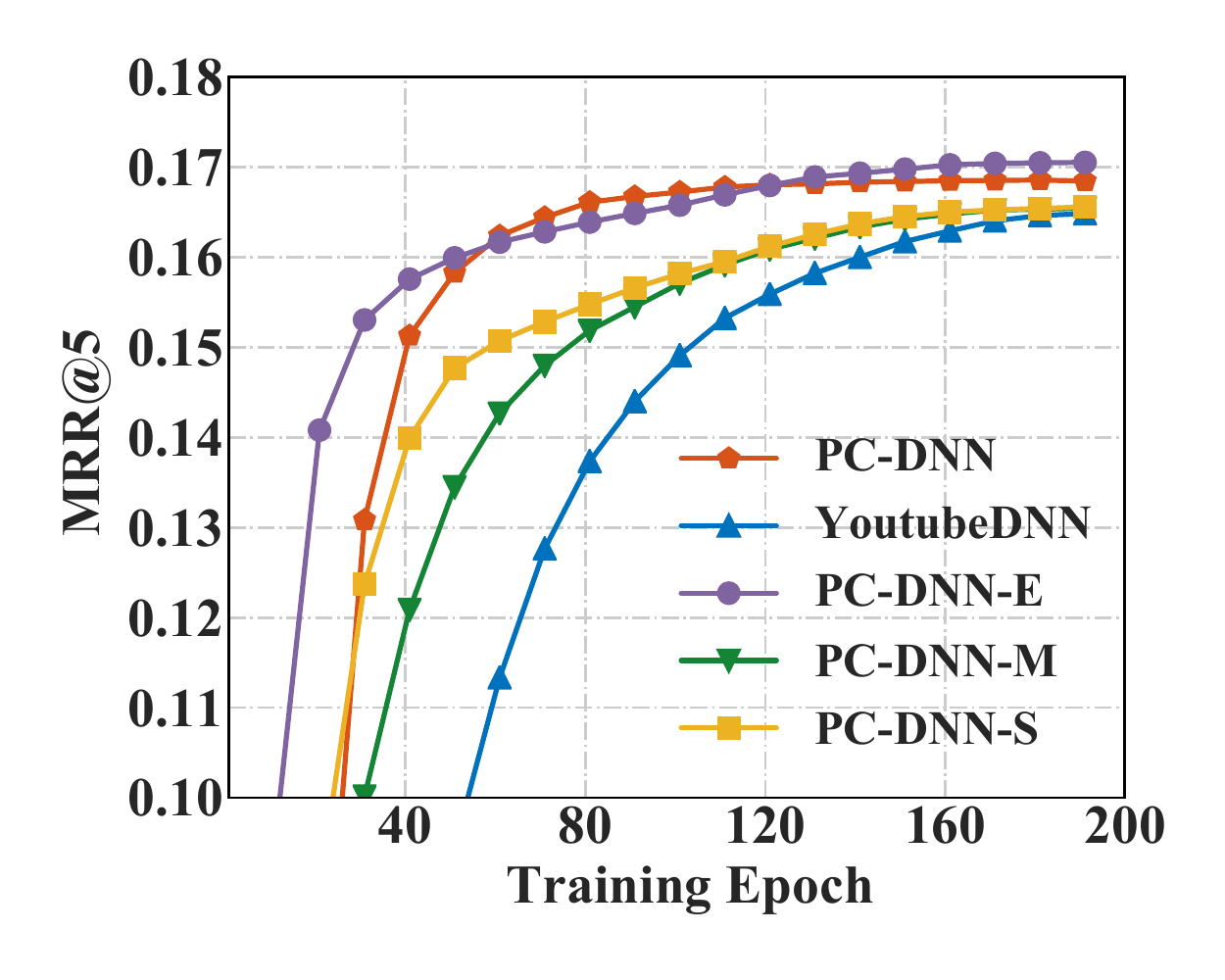}
            \subcaption{PC-DNN on Retailrocket}
    \end{subfigure}
    \begin{subfigure}[t]{0.24\linewidth}
    \centering
            \includegraphics[width=1.7in]{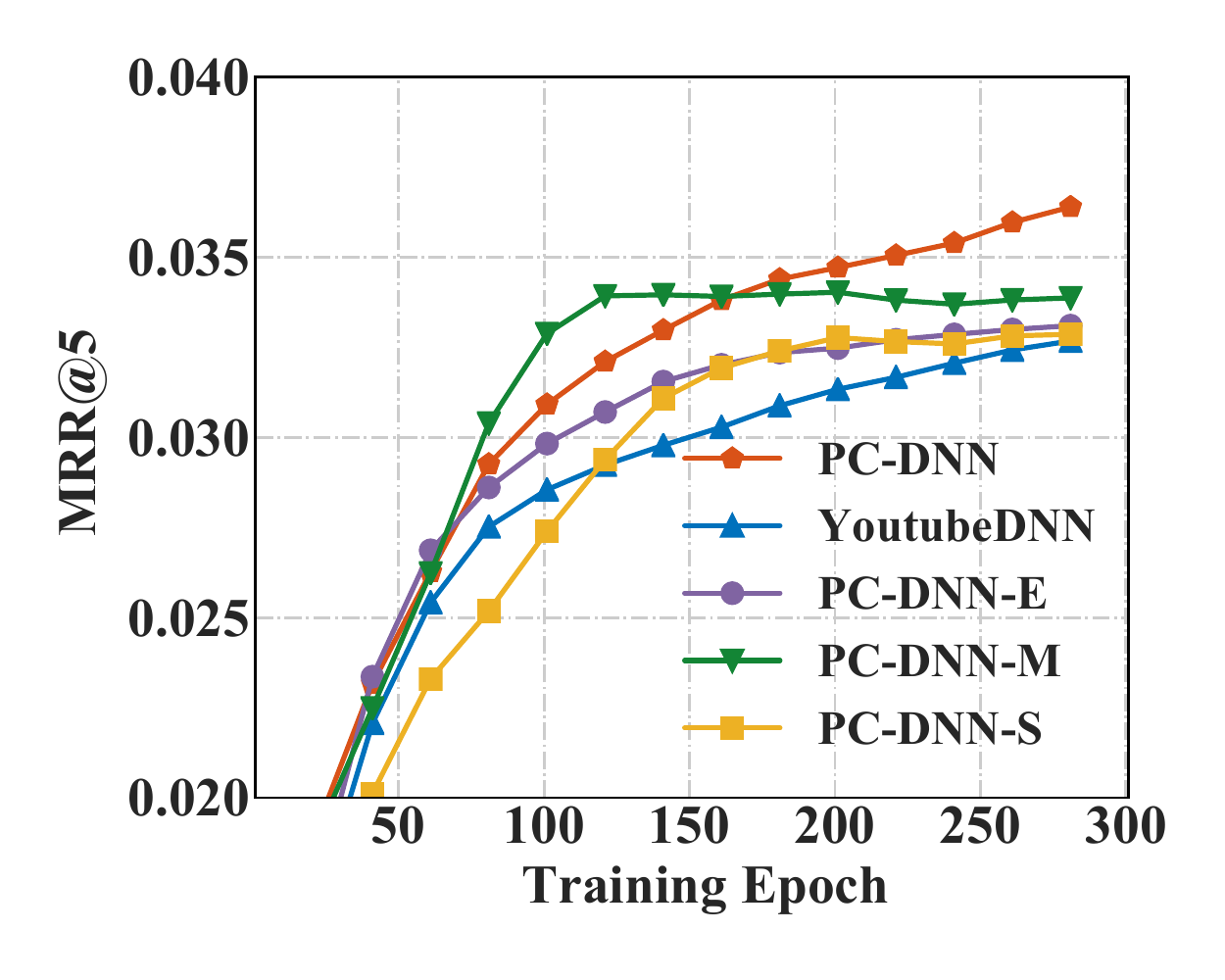}
            \subcaption{PC-DNN on ML-20M}
    \end{subfigure}
    \caption{Convergence behaviors of PCRec by peer collaboration of different components. PC-SAS that applies peer collaboration only on the embedding layer, middle layers, and softmax layer as {PC-SAS-E}, {PC-SAS-M}, {PC-SAS-S}, respectively. 
    % and {SeCo-NFM-E}, {SeCo-NFM-M} and {SeCo-NFM-S}, respectively.  
    % Among them, we use two variants of SeCo-Next-E and SeCo-NFM-E about the input layer, namely, SeCo-Next-EO that make the input layer integrally for self-cooperation method and PCRec-N-EV that make each embedding vector of the embedding layer for information complement.
    }
    \label{compontent}
\end{figure*}

\subsection{Ablation Study (RQ3,4)} 

% \subsubsection{Impact of measure unit (RQ2)}
% We consider two ways to complement information in PCRec in Section~\ref{citeria} which is layer and invalid (single) weights as measure unit and examine the two ways in Table~\ref{measureunittable}. As shown, SeCo-Next-L always yields better performance compared with SeCo-Next-I, which proves fully considering the network connectivity is important. Surprisingly, SeCo-Next-I which only focus on invalid weights outperforms the original NextItNet. We observe similar results for SeCo-NFM. These results suggest that PCRec should make layer as measure unit which fully considering the network connectivity help the valid weights to be reactivated.

\subsubsection{Impact of learning rates and train data orders}
% Consider the model structure, NextItNet can capture sequential dependency of user-item interactions and the best recommendation accuracy by stacking dilation convolution layer, YouTubeDNN obtain interaction pattern of user-item by stacking full connection layer that exist in NeuralFM. And NeuralFM model the non-linearly high-order feature interactions of user-item by combining low-order interactions with full connection module, which is fundamentally different from YouTubeDNN. Besides,  NeuralFM and YouTubeDNN are belong to collaborative filtering model. We argue NeuralFM for further study is more value than YouTubeDNN. Along with presenting all results on the four datasets is redundant and space unacceptable. Hence, in the rest of this paper, we focus on comparing our proposed improvements with NextItNet and NeuralFM and report parts of the follow-up results with respect to PCRec on some datasets and metrics considering that their behaviors are consistent. sampling order

Table~\ref{rateandorder} presents the impact of different learning rates and sampling orders of training data. As shown, PC-SAS-DD  always yields the best recommendation accuracy compared with their counterparts, i.e.,PC-SAS-SD and PC-SAS-DS. On the other hand, we observe that PC-SAS-SD and PC-SAS-DS consistently outperform the original SASRec. Similar observations can be made for PC-DNN. The results confirm that PCRec that applies different learning rates and sampling orders is necessary. This is likely because training individual networks with different learning rates and sampling could increase diversity of network weights, so as to increase the layer information when linearly combining them. The results hold well for PCRec with BPR and are thus simply omitted.

% the optimal performance of PCRec should
% is obtained with different learning rates and orders of training data. 

% One explanation for the advantage of PCRec is that parameters of multiple models are diverse and 
% for to verify using different learning rate and train data order enhancing diversity and information of multiple models. 
%  the results indicate that SeCo-Next-DD always yields the best recommendation accuracy on all datasets. In addition, SeCo-Next-SD and SeCo-Next-DS have achieved better performance than NextItNet, which indicates the two strategies have a contribution for the cooperative process of multiple models. Among these strategies, SeCo-Next-DS performs better than SeCo-Next-SD. These results suggest that the different learning rates can further promote the effect of information complement than the different train data order. Thus, in subsequent experiments, we always use the two strategies as a default setup.

\subsubsection{Impact of $\alpha$}
In this subsection, we study the impact of $\alpha$ which controls the amount of information to be transplanted. Figure~\ref{hyper-alpha} shows the model performance of PC-SAS and PC-DNN with different $\alpha$ on Retailrocket and ML-20M. First, PC-SAS is sensitive to $\alpha$, and the optimal results are obtained when $\alpha$ equals to 30 on Retailrocket and ML-20M.
Similarly, PC-DNN obtains the best performance when $\alpha$ is set to 40 on Retailrocket and ML-20M. It can be seen that PCRec with a proper $\alpha$ could achieve 1$\sim $4\% improvement than a random $\alpha$.
% On the other hand, SeCo-NFM show similar behaviors on Weishi, but is insensitive to $\alpha$ on ML30. 
It is also worth noting that PCRec outperforms its individual base model even $\alpha$ is not set to its optimal value. In practice, we suggest running PCRec by tuning $\alpha$ from 30 to 40. By doing this, we observe that the coefficient $\mu_k$ ranges from 0.7 to 1.0 in most cases.
\subsubsection{Impact of the peer collaboration with different components}
We conduct an ablation study in Figure~\ref{compontent} by applying peer collaboration for some components of the model.
% We denote {PC-SAS} that apply peer collaboration only on the embedding layer, hidden layers, and softmax layer as {PC-SAS-E}, {PC-SAS-M}, {PC-SAS-S}, respectively. 
% Here, we omit the results of PC-BPR because it only has embedding layer and show similar behaviors. 
First, it can be observed that PC-SAS-E, PC-SAS-M and PC-SAS-S outperform SASRec, demonstrating that the information transplanting on every component of usually performs better than its original model SASRec. Second, PC-SAS-E improves SASRec by a larger margin, compared with PC-SAS-M. In particular, PC-SAS-E even surpasses PC-SAS on ML-20M.  This is likely because the embedding layer usually contains much more parameters than the middle layers in recommender models. Besides, the embedding layer contains the most important information for item recommendations --- i.e., personalization. As such, performing information transplanting on the embedding layer makes more sense than only doing it for these middle layers.
% We argue that the embedding matrix takes up most parameters of model() and peer-collaboration helps with item representation learning, which is the reason why PC-BPR obtains a satisfying results. 
This also suggests that it might be sufficient to perform peer collaboration on only necessary components of the base model, rather than all components.
% {In particular, PC-SAS-E and PC-SAS-S improve NextItNet by a larger margin compared with PC-SAS-M. It is reasonable since the embedding and softmax matrices have more parameters than these hidden layers. Meanwhile, PC-SAS shows better results than PC-SAS-E, PC-SAS-M and PC-SAS-S. This also suggests that it is meaningful to perform self-cooperation on all components of the base model.} 
Similar conclusions hold for PC-DNN in general.

\begin{figure}[t]
	\centering
	\begin{subfigure}[t]{0.23\textwidth}
		\centering
		\includegraphics[width=1.6in]{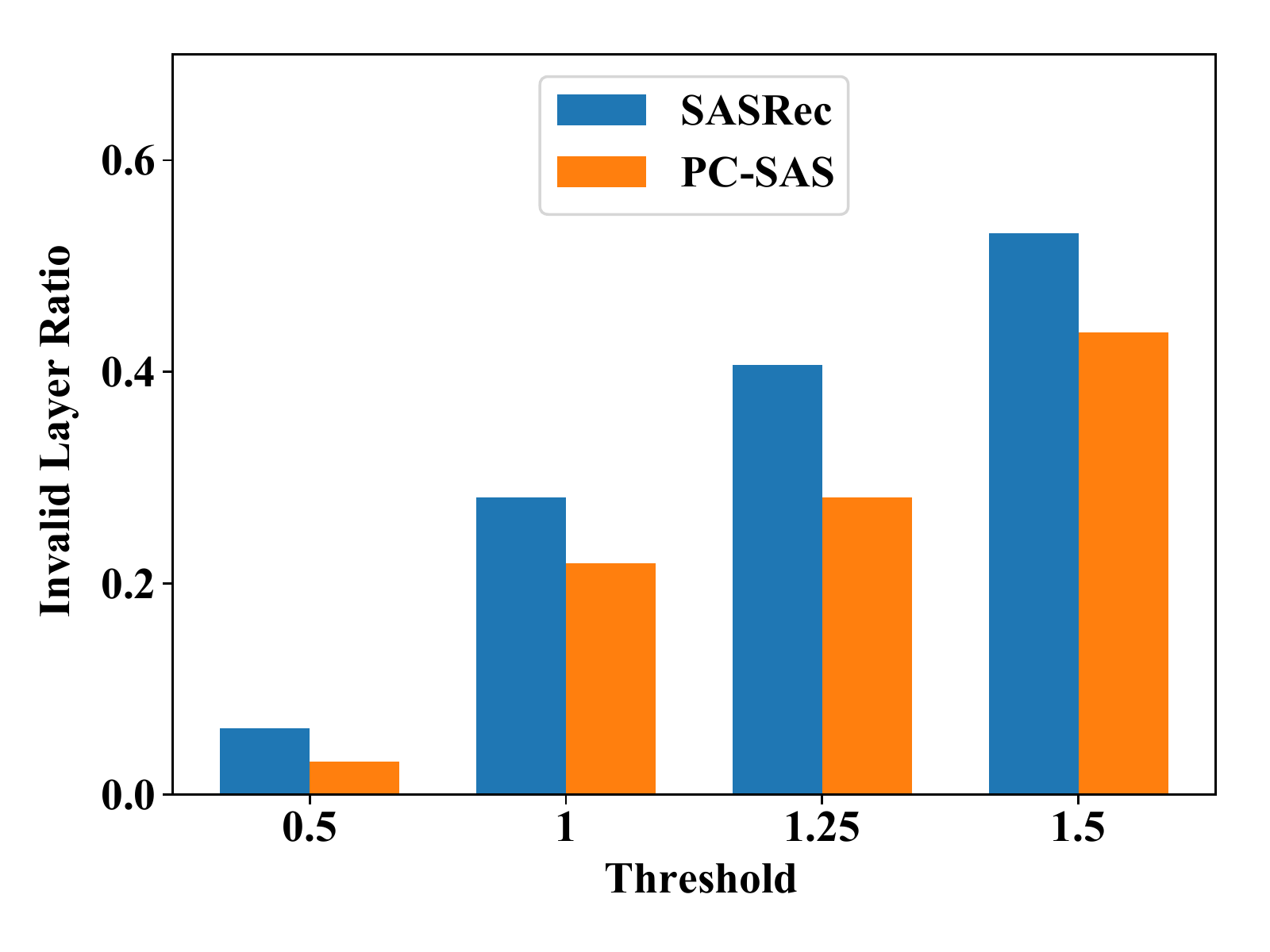}
		\subcaption{Retailrocket}
		\label{pw}
	\end{subfigure}%
	\begin{subfigure}[t]{0.23\textwidth}
		\centering
		\includegraphics[width=1.6in]{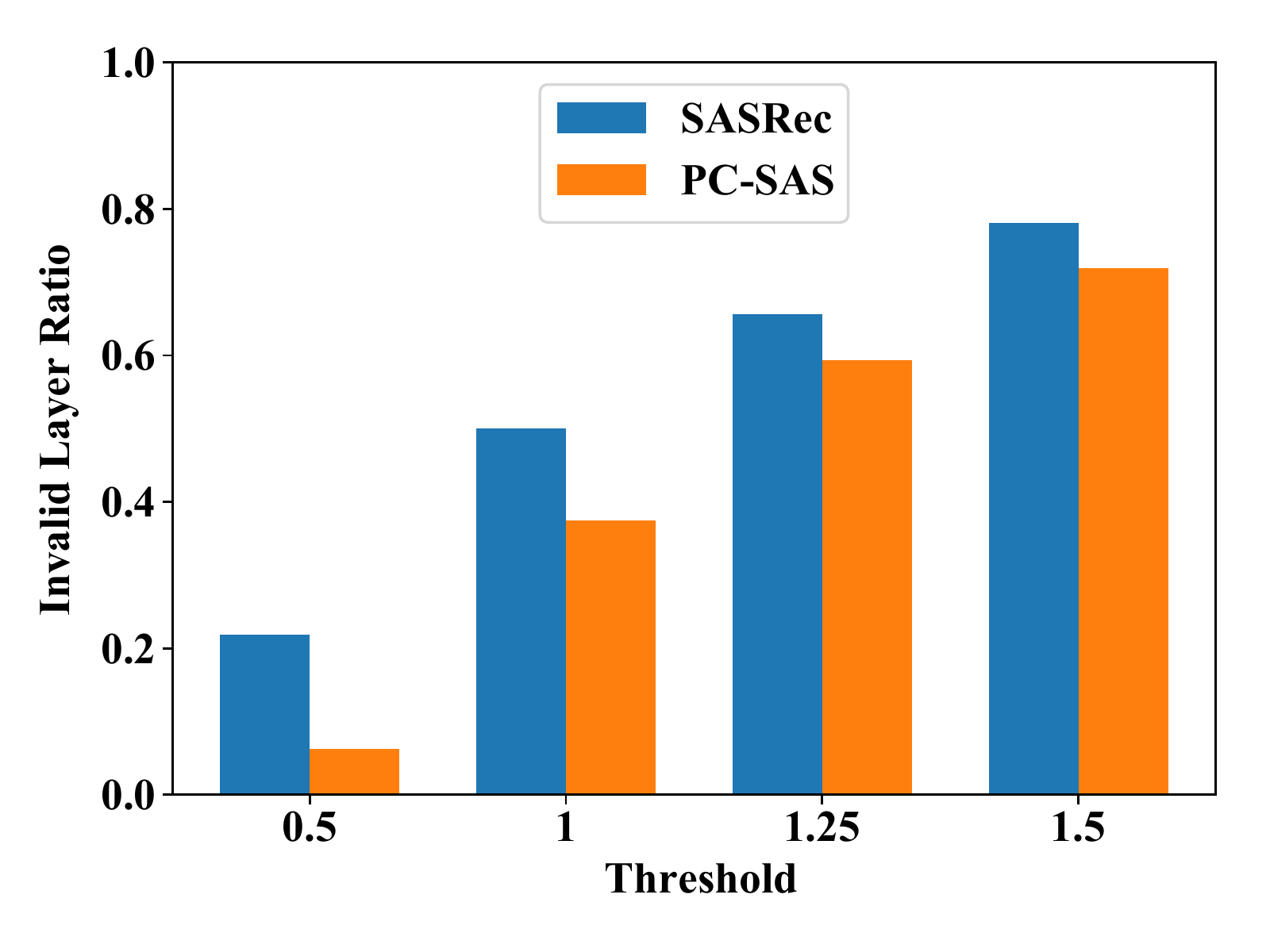}
		\subcaption{ML-20M}
		\label{pm}
	\end{subfigure}
	\caption{Ratios of Invalid layers in PC-SAS and SASRec, where invalid layer ratio denotes the number of valid layers (with information $H(W_{i})$ under a specified threshold value) over the number of all layers.}
	\label{invalidratio}
\end{figure}

\subsection{Effectiveness of PCRec (RQ5)}
In this part, we simply analyze the information transplanting mechanism in PCRec. To validate whether the peer collaboration really work not, we calculate the number of invalid layers (including both fully-connected layers and self-attention layers) whose entropy is under a specified threshold after training. Experimental results are reported in Figure~\ref{invalidratio}. It can be seen that with the threshold of 0.5, there are about 20\% layers that are invalid for SASRec on ML-20M, whereas PCRec with peer collaboration training only has less than 5\% invalid layers. With the increase of threshold, the ratios of invalid layers in both SASRec
and PCRec rise. However, the ratio of PCRec is always smaller than SASRec. These observations verify our key assumption that peer collaboration does help model to strengthen the information (i.e., $H(W_{i})$) of network layers.

\section{CONCLUSION}
In this work, we have discussed the network redundancy phenomenon in deep recommender models. Taken inspiration from this, we have proposed PCRec, a flexible and generic peer collaboration learning paradigm that is able to rejuvenate invalid parameters (instead of abandoning them) in a recommender model by transplanting information from its outside peer network. 
To identify which parameters are invalid, we have introduced L1-norm and entropy based criteria. Then, we propose two collaboration strategies regarding how to transplant information between two peer models. Through extensive experiments on three real-world recommendation datasets,  we have demonstrated that PCRec generated consistently better recommendations than its original base model. We expect PCRec to be valuable for existing recommender systems based on the embedding or deep neural network models.

% we argued existing over-parameterized issue in RS, and performed empirical studies to justify this argument. Then we propose PCRec, a flexible \& generic multiple models self-cooperation framework for leveraging these unimportant weights of model, which consisting of four essential condition. First of all, we make the layer of model as measure unit. And we adapt entropy as criterion to evaluate the information (unimportant weights) of each layer. Then we proposed adaptive coefficient strategy and linear combination method to complement information between each layer of the multiple models. Through extensive experiments on real-world datasets, we demonstrate that PCRec generates recommendations with higher accuracy and better generalization ability. An important conclusion made from these results is that the commonly used recommender models have over-parameterized at all. 

\bibliographystyle{ACM-Reference-Format}
\bibliography{sample-base}

\end{document}